\documentclass[12pt,a4paper]{article}
\usepackage{graphicx}
\usepackage{subeqnarray}
\begin{document}
%
%
%
%
\def\astrobj#1{#1}
\newenvironment{lefteqnarray}{\arraycolsep=0pt\begin{eqnarray}}
{\end{eqnarray}\protect\aftergroup\ignorespaces}
\newenvironment{lefteqnarray*}{\arraycolsep=0pt\begin{eqnarray*}}
{\end{eqnarray*}\protect\aftergroup\ignorespaces}
\newenvironment{leftsubeqnarray}{\arraycolsep=0pt\begin{subeqnarray}}
{\end{subeqnarray}\protect\aftergroup\ignorespaces}
\newcommand{\diff}{{\rm\,d}}
\newcommand{\pprime}{{\prime\prime}}
\newcommand{\szeta}{\mskip 3mu /\mskip-10mu \zeta}
\newcommand{\FC}{\mskip 0mu {\rm F}\mskip-10mu{\rm C}}
\newcommand{\appleq}{\stackrel{<}{\sim}}
\newcommand{\appgeq}{\stackrel{>}{\sim}}
\newcommand{\Int}{\mathop{\rm Int}\nolimits}
\newcommand{\Nint}{\mathop{\rm Nint}\nolimits}
\newcommand{\range}{{\rm -}}
\newcommand{\displayfrac}[2]{\frac{\displaystyle #1}{\displaystyle #2}}
\def\astrobj#1{#1}
%
\title{A simple multistage closed-(box+reservoir) model
of chemical evolution}
\author{{R.~Caimmi}\footnote{
{\it Astronomy Department, Padua Univ., Vicolo Osservatorio 3/2,
I-35122 Padova, Italy}
email: roberto.caimmi@unipd.it~~~
fax: 39-049-8278212}
\phantom{agga}}
%
%
\maketitle
\begin{quotation}
\section*{}
\begin{Large}
\begin{center}

Abstract

\end{center}
\end{Large}
\begin{small}

\noindent\noindent
Simple closed-box (CB) models of chemical evolution are
extended on two respects, namely (i) simple closed-(box+reservoir)
(CBR) models allowing gas outflow from the box into the
reservoir (Hartwick 1976) or gas inflow into the box
from the reservoir (Caimmi 2007) with same composition
as the preexisting gas and rate proportional to the
star formation rate, and (ii) simple multistage
closed-(box+reservoir) (MCBR) models allowing different
stages of evolution characterized by different inflow
or outflow rates.   In any case, the stellar initial mass
function is assumed to be universal, and mass conservation
holds for the whole system (box+reservoir) while it is
violated for each subsystem (box and reservoir).   The
theoretical differential oxygen abundance distribution
(TDOD) predicted by the model, under the assumption of
instantaneous recycling, is a continuous broken line,
where outflow and moderate inflow rates are represented
by negative slopes, steady inflow rates by null slopes,
and strong inflow rates by positive slopes.   Then an
application is made to a special stellar system
resembling the inner Galactic halo.   To
this aim (a) a fictitious sample is built up from two
distinct samples (Ryan and Norris 1991; Sch\"orck et al.
2009) and taken as representative of the inner Galactic
halo, and (b) different [O/H]-[Fe/H] empirical relations
are deduced from different samples (Rich and Boesgaard
2009; Fabbian et al. 2009; Schmidt et al. 2009) related
to different methods, and two of them are selected for
determining the empirical differential oxygen abundance
distribution (EDOD) with regard to the fictitious sample.
In both cases the EDOD is represented, to an acceptable
extent, as a continuous broken line.   The slopes and the
intercepts of the regression lines are determined, and
then used as input parameters to MCBR models with fiducial
values assigned to the remaining input parameters.
Output parameters are the gas mass fraction and the star
mass fraction at the end of each evolutionary stage.
If the inner halo and the metal-poor bulge (after the inflow stage)
are represented by the box and the reservoir, respectively,
then the inner halo fractional mass (normalized to the halo)
is comparable with, or exceeds by a factor up to 4, the
metal-poor bulge fractional mass (normalized to the bulge),
for current estimates of the halo-to-bulge mass ratio of
about 0.05-0.10.   On the other hand, quantitative results
cannot be considered for applications to the inner Galactic 
halo, unless
selection effects and disk contamination are removed
from halo samples, and discrepancies between different
oxygen abundance determination methods are explained.

\noindent
{\it keywords - 
galaxies: evolution - stars: formation; evolution.}
\end{small}
\end{quotation}

\section{Introduction} \label{intro}

The empirical metallicity distributions of long-lived
stars of different populations, constrain models for
the formation and the evolution of the Galaxy.  Simple
closed-box (CB) models make a useful tool in the
description of galactic chemical evolution.   The
original formulation (Searle and Sargent 1972; Pagel
and Patchett 1975) relies on instantaneous recycling
and homogeneous mixing approximations i.e. gas from
dying stars is instantaneously returned to and
homogeneously mixed with the interstellar medium.
In addition, (1) a null star mass fraction and a null
metal abundance are taken as initial values, and (2)
mass conservation is assumed to hold, which implies
no gas outflow from the box or inflow into the box.

Simple CB models may be extended in many respects,
such as nonzero initial metal abundance (prompt
initial enrichment) and/or star mass fraction
(Truran and Cameron 1971), inhomogeneous mixing
(Caimmi 2000 + erratum 2001a, hereafter quoted
together as C00; Caimmi 2001b, hereafter quoted
as C01), and gas outflow (Hartwick 1976, hereafter
quoted as H76) or inflow (Caimmi 2007, hereafter
quoted as C07) with same composition as the
preexisting gas and rate proportional to the star
formation rate.    Formulations implying more
extreme changes, such as reject of instantaneous
recycling and unconditioned gas outflow or inflow,
can no longer be considered as ``simple'' models.

The existence of a G-dwarf problem i.e. detection
of too few metal deficient G dwarf (or, more
generally, of a selected spectral type) with
respect to that which could be expected from
simple CB models of chemical evolution (e.g.,
Searle and Sargent 1972; Pagel and Patchett
1975; Haywood 2001) was first established in
the solar neighbourhood (van den Bergh 1962;
Schmidt 1963).   Although in a less extreme
form, a G-dwarf problem appears to exist both
in the halo (e.g., H76; Prantzos 2003) and in
the bulge (e.g., Ferreras et al. 2003).   In
addition, a G-dwarf problem has been recognized
in both bulge-dominated and disk-dominated
galaxies (Henry and Worthey 1999), which is
consistent with the idea that the G-dwarf
problem is universal (Worthey et al. 1996).

According to current $\Lambda$CDM cosmological
scenarios, galaxies were largely built out of
disrupted smaller subunits (dSph galaxies whose
surviving cores could be massive globular clusters).
A similar process is presently occurring on a
larger scale: a central bulge (a cD galaxy) is
accreting in galaxy clusters, at the expense
of infalling smaller galaxies.   In this view,
a universal G-dwarf problem appears to be
closely related to the initial assembling
stage of galactic evolution.

With regard to the Galaxy, the empirical
differential oxygen abundance distribution
(EDOD) shows that (i) at least two different
stages of (chemical) evolution exist, and
(ii) each stage can be described using an
extended simple CB model, for both the inner halo
(C07), the bulge (C07), the thick disk
(Caimmi and Milanese 2009, hereafter quoted
as CM09), and the thin disk (CM09).   In
particular, the earlier stage is marked
by an initially increasing EDOD.

The
advantage of the EDOD, $\psi=\log[\Delta N/
(N\Delta\phi)]$%
\footnote{Caption of symbols: $\phi=Z_{\rm O}/
(Z_{\rm O})_\odot$ is the oxygen abundance
normalized to the solar value, $N$ is the population
of the sample under consideration, and $\Delta N$
is the number of sample objects within a bin centered
in $\phi$ and bounded at $\phi\mp\Delta\phi/2$.},
on the oxygen abundance
distribution, $\Psi=\Delta N/N$, is that
the theoretical differential oxygen abundance
distribution (TDOD) predicted by extended
simple CB models is a straight line on the
$({\sf O}\phi\psi)$ plane, which provides
a more sensitive test (Pagel 1989; Malinie
et al. 1993; Rocha-Pinto and Maciel 1996;
C00; C01; C07; Caimmi 2008; CM09).

As shown in recent attempts (Carollo et al.
2007, 2010), the Galactic halo may be divided into
two structural components, namely (1) an
inner spheroid with axes, $a\approx15\,$kpc,
$c\approx10\,$kpc, which exhibits a modest
prograde rotation and a metallicity peak
at [Fe/H] $\approx-1.6$, and (2) an outer
sphere with radius, $R\approx20\,$kpc, which
exhibits a clear retrograde net rotation
and a metallicity peak at [Fe/H] $\approx-2.2$.
In addition, the inner halo population
dominates within its own volume, while
the outer halo population includes a
larger fraction of low-metallicity
([Fe/H] $<-2.0$) stars than does the
inner halo population.

For a component of orbital motion
measured with respect to the Local
Standard of Rest, $V<200\,{\rm km\,s^{-1}}$,
the distribution of [Fe/H]
appears similar to what in the past
was considered ``the halo'', with a
single metallicity peak at [Fe/H]
$\approx-1.6$ (Carollo et al. 2007,
Fig.\,2), which implies the inner
halo population is dominant in the
related sample.   A similar trend
is exhibited by the metallicity
distribution inferred from an earlier
attempt (Ryan and Norris 1991, hereafter
quoted as RN91, Fig.\,5c), and the
related sample (hereafter quoted as the
RN91 sample) can
also be considered as representative
of the inner halo.  In both cases,
data are biased towards low metallicities,
[Fe/H] $<-3.0$.

The inner halo low-metallicity tail has
been determined using data from the Hamburg/ESO
survey (Sch\"orck et al. 2009, hereafter quoted
as H\,V, Fig.\,11), but the related sample
(hereafter quoted as the H\,V sample) is
biased towards higher metallicities, [Fe/H]
$>-3.0$, and extremely low metallicities,
[Fe/H] $<-4.2$.   In the comparison of the
empirical with the theoretical metallicity
distribution, selection effects must be
taken into account.   Accordingly, each
theoretical metallicity distribution under
consideration has to be converted into
its counterpart as it would be observed
by applying the metal-poor star selection
criteria used in the survey (H\,V).

If, on the other hand, the H\,V sample
is dominated by the inner halo population,
a correction for the above mentioned
selection effects is expected to yield a
metallicity distribution with a single
peak at [Fe/H] $\approx-1.6$, similar
to its counterpart related to earlier
samples (RN91; Carollo et al. 2007, Fig.\,2,
$V<200\,{\rm km\,s^{-1}}$).   In
fact, the metallicity distributions
related to the RN91 and H\,V sample
can be adjusted to match in the range,
$-3.4<{\rm [Fe/H]}<-2.6$, where selection
effects are negligible (H\,V, Figs.\,11, 12).

The Milky Way inner halo offers a
unique opportunity to construct the
star formation and mass assembly
history of a galactic inner halo,
hence providing a unique benchmark
for theories of galaxy formation
and evolution.   To this aim, the
sample used has to be complete at
least for a wide range of metallicity.
In the case under discussion,
$-3.6<{\rm [Fe/H]}<-2.8$ for the H\,V sample and
$-3.4<{\rm [Fe/H]}<-1.0$ for the RN91 sample.
The remaining ranges are affected by
various biases, mainly due to the
occurrence of selection effects for
the low-metallicity tail above a
threshold (H\,V)
and disk contamination for the
high-metallicity tail (RN91, H\,V).

The metallicity distribution of the
inner halo can be deduced from a
fictitious sample (hereafter quoted
as the fs10 sample) within the range,
$-4.2<[$Fe/H$]<+0.2$, under a number of
restrictive but reasonable assumptions,
namely (i) the H\,V sample is
representative of the inner halo
within the range, $-4.2<[$Fe/H$]<-3.0$;
(ii) the RN91 sample is representative
of the inner halo within the range,
$-2.8<[$Fe/H$]<+0.2$; (iii) the H\,V
and RN91 samples are equally representative
(even if related to different populations)
of the inner halo within the range,
$-3.0\le[$Fe/H$]\le-2.8$.

In absence of direct [O/H] evaluations,
the EDOD related to the fs10 sample can
be deduced by use of a [O/H]-[Fe/H]
empirical relation which, on the other
hand, is strongly dependent on both
the selection of spectroscopic oxygen
lines and the choice of the model, as
shown in recent attempts (Ramirez et al.
2007; Rich and Boesgaard 2009, hereafter
quoted as RB09; Fabbian et al. 2009,
hereafter quoted as Fal09; Schmidt et al.
2009, hereafter quoted as Sal09) where
earlier attempts are quoted (for further
insight refer to the proceedings edited
by Barbuy et al. 2001).   The discrepancy
between [O/H]-[Fe/H] empirical relations,
deduced using different methods and
different models, still remains large.
For this reason, the TDOD should be
fitted to the EDOD related to two
different [O/H]-[Fe/H] empirical
relations, as done in previous papers
(C01; C07; Caimmi 2008; CM09).

Simple CB models of chemical evolution
were first extended allowing for gas
outflow (H76) and later for moderate gas
inflow, with same metal abundance as in
the preexisting gas, yielding negative
TDOD slopes in the $({\sf O}\phi\psi)$
plane.   On the other hand, the
low-metallicity tail of the EDOD is
fitted to a straight line with positive
slope for both the halo (C07), the
bulge (C07), the thick disk (CM09),
and the thin disk (CM09), which is
achieved by extending simple CB models
to strong gas inflow with same metal
abundance as the preexisiting gas.

In this view, different stages of evolution
are related to different domains in the
normalized oxygen abundance, $(\phi_{\rm U})_i
\le\phi\le(\phi_{\rm U})_f$, where the EDOD
is fitted to a regression line, $\psi=a_{\rm U}
\phi+b_{\rm U}$, with slope, $a_{\rm U}$, and
intercept, $b_{\rm U}$ (${\rm U}={\rm I},
{\rm II},...,$ is the stage
considered, beginning and ending at $i$ and $f$
configurations, respectively).   For reasons of
continuity, the final values of parameters
related to a selected stage must necessarily
coincide with the corresponding initial values
related to the next stage, with the exception
of the outflow or inflow rate, which can be deduced
from the EDOD.  Accordingly, the knowledge of
present values allows the calculation of
past values, in the light of the model, getting
insight on the formation and evolution of the
system under consideration.

In summary, the current paper is aimed to the
following main points: (i) metallicity distribution
related to a fictitious sample (fs10), supposed to
be representative of the inner halo (subsection
\ref{fs10}); (ii) choice of two different
[O/H]-[Fe/H] empirical relations deduced from recent
samples (RB09, Fal 09, Sal09) (subsection \ref{OFre});
(iii) EDOD determination from the fs10 sample by use
of the above mentioned [O/H]-[Fe/H] empirical relations
(subsection \ref{EDOD}); (iv) formulation of extended
simple CB models, namely (a) simple CBR models, allowing
for gas outflow or inflow with same metal abundance
as the preexisting gas, and (b) simple MCBR models,
allowing for different stages of evolution
(subsections \ref{bath}, \ref{TDOD} and \ref{mscm});
(v) determination
of the best fitting TDOD to the EDOD inferred from the
fs10 sample (subsection \ref{FEDO}); (vi) application
to a special stellar system resembling the inner
Galactic halo (subsection \ref{feih}).
In addition, the discussion and the conclusion make
the subject of subsection \ref{disc} and section
\ref{conc}, respectively.

\section{Data and inferred quantities} \label{data}

The following ingredients are necessary for the
determination of the EDOD (in particular, related to the
inner Galactic halo), from which input parameters of a
selected model are deduced: (i) a representative sample;
(ii) a related [Fe/H] distribution; (iii) a [O/H]-[Fe/H]
empirical relation.   Each point shall be separately
dealt with in the forthcoming subsections.

\subsection{Building up a fictitious sample (fs10)} \label{fs10}

In building up a fictitious sample (the fs10 sample) of
inner halo stars, two different samples shall be considered,
namely the RN91 sample ($N=372$) of kinematically selected
halo subdwarfs and the H\,V sample ($N=3439$) of metal-poor
stars selected from the Hamburg/ESO objective prism survey.
More specifically, the H\,V sample has been determined
from a subsample ($N=1638$) with available spectroscopic
follow-up observations, by means of scaling to the full
parent sample ($N=3439$).   For further details refer to
the parent paper (H\,V).
The RN91 sample can be considered as complete within the
range, $-3.4<[$Fe/H$]<-1.0$.   At different abundance the
sample is incomplete and suffers contamination from disk
stars at the high-metallicity tail, [Fe/H$]\ge-1.0$.
The H\,V sample can be considered as complete within the
range, $-4.2<[$Fe/H$]<-2.8$.   At different abundance the
sample is incomplete and suffers both selection effects
([Fe/H$]>-2.8$) and contamination from disk stars ([Fe/H$]>
-2.0$).

Contamination from outer halo stars also exists
in both samples, but the effect is expected to be negligible.
In fact, the inner halo dominates within a spheroid centered
on the Galactic centre, with equatorial plane coinciding
with the Galactic plane, and semiaxes $(a,c)=(15,10)$
kpc (Carollo et al. 2007).   On the other hand, both the
RN91 and H\,V sample are made mainly of stars placed
within the above mentioned spheroid.   Then the contamination
from outer halo stars is expected to be appreciable only
at low ([Fe/H$]<-2.0$) metal abundances (Carollo et al. 2007).

The metallicity distribution from the RN91 sample
peaks at [Fe/H$]\approx-1.6$, in agreement with
recent results for the inner halo population
(Carollo et al. 2007, 2010).   On the contrary, the
metallicity distribution related to the H\,V
sample appears to be bimodal with an absolute
maximum at [Fe/H$]\approx-2.2$ and a relative
maximum at [Fe/H$]\approx-1.0$.   The bimodality
is probably an artefact due to selection effects
and/or disk contamination for the following
reasons%
\footnote{The problem was raised via e-mail
(December 2009) to two coauthors of the parent
paper (H\,V) and to two coauthors of the paper
where inner and outer halo populations were first
detected (Carollo et al. 2007), but no reply was
received up to-day.  It cannot be
excluded that related messages have been lost
into the spam bin.}:
(i) the H\,V sample is mainly made of inner
halo stars (H\,V); (ii) the expected peak at
[Fe/H$]\approx-1.6$ (Carollo et al. 2007, 2010) is
lacking; (iii) the metallicity distribution
from the RN91 sample can be scaled to match
its counterpart from the H\,V sample within
the range, $-3.4<[$Fe/H$]<-2.6$ (H\,V);
(iv) a bimodal metallicity distribution cannot
be fitted to its theoretical counterpart
deduced from simple models of chemical evolution,
even if the latter is ``observed'' by applying
the metal-poor selection criteria used in the
Hamburg/ESO survey (H\,V).

The H\,V sample is mainly made of giant stars.
A subsequent study on the stellar content of
the Hamburg/ESO survey has been focused on a
sample $(N=617)$ of main sequence turnoff stars
(Li et al. 2010).   Both samples exhibit a quite
analogous metal abundance distribution where the
halo population dominates ([Fe/H$]<-2.0$), while
the contrary holds for higher values ([Fe/H$]>-2.0$).
In the latter case, the H\,V sample is also affected
from the survey-volume effect.   For further details
refer to the parent paper (Li et al. 2010).

Basing on the above considerations, the following
working hypotheses are made.
\begin{description}
\item[(1)]
The H\,V sample $(N=3439)$ is representative
of the inner halo within the range, $-4.2<[$
Fe/H$]<-2.8$, where selection effects are
minimized and contamination from disk stars
is negligible (H\,V).
\item[(2)]
The RN91 sample $(N=372)$ is representative
of the inner halo within the range, $-3.0<[$
Fe/H]$<+0.2$, with a caveat due to contamination
from disk stars for [Fe/H$]>-1.0$ (RN91).
\item[(3)]
The RN91 and H\,V sample are equally representative
(even if belonging to different populations) of
the inner halo within the range, $-3.0\le
[$Fe/H$]\le-2.8$, where the number of stars
is $(\Delta N)_{\rm RN91}=8$ and $(\Delta N)_
{\rm H\,V}=160$, respectively.
\end{description}
Accordingly, the fs10 sample can be built up
from the RN91 and H\,V samples, by normalizing
to the same number of stars within the metallicity
range where the above mentioned samples are
supposed to be equally representative of the
inner halo.   More specifically, the normalization
shall be performed with respect to the H\,V
sample, but the Poissonian error related to each
bin shall remain unchanged with respect to the
parent sample, with the minimum among the two
retained for the bin where the parent samples are
equally representative of the inner halo.
Then the number of stars belonging to the fs10
sample, with regard to a selected metallicity bin,
remains unchanged or is multiplied by a factor 20
according if the parent sample is H\,V ([Fe/H$]
\le-2.8$) or RN91 ([Fe/H$]>-2.8$), yielding a
fictitious population, $N=7452$.

\subsection{The [O/H]-[Fe/H] empirical relation} \label{OFre}

In dealing with simple models of chemical
evolution, involving the assumption of
instantaneous recycling, the predicted
metal abundance has to be compared to
the observed oxygen (or any other primary
element) abundance (e.g., Pagel 1989; C00;
C01).   Unfortunately, oxygen is more
difficult than iron to detect, and an
empirical formula may be needed to
express the former as a function of the
latter.   The population of available
samples where oxygen abundances are directly
determined, does not exceed a few hundreds
at most (e.g., Ramirez et al. 2007;
Melendez et al. 2008; RB09; Fal09; Sal09).
With regard to the halo, only the RB09 and
Fal09 samples can be considered as representative,
while the Sal09 sample shall be included for
comparison, and the remaining above quoted two
shall be excluded.

The RB09 sample $(N=49)$ is made of a homogeneous
subsample $(N=24)$ of metal-poor ($-3.5<[$Fe/H$]<
-2.2$) stars, and a non homogeneous subsample
$(N=25)$ of higher-metallicity ($-3.1<[$Fe/H$]<
-0.5$) stars.   In both cases, the stellar
population remains unspecified and oxygen abundance
has been determined using standard local thermodynamical
equilibrium (LTE) one-dimensional hydrostatic model
atmospheres.   The calculated oxygen abundance is
independent of the solar value.   Typical uncertainties
are $\Delta[$Fe/H$]=\mp0.15$ and $\Delta[$O/H$]=\mp0.15$.
Standard deviations are also provided for each star.
For further details refer to the parent paper (RB09).

The Fal09 sample $(N=43)$ is made of halo stars ($-3.3<
[$Fe/H$]<-1.0$) where oxygen abundance has been determined
using three different methods involving (a) LTE one-dimensional
hydrostatic model atmospheres; (b) three-dimensional
hydrostatic model atmospheres in absence of LTE with
no account taken of the inelastic collisions via neutral
H atoms $(S_{\rm H}=0)$, hereafter quoted as SH0; (c)
three-dimensional hydrostatic model atmospheres in absence
of LTE with due account taken of the inelastic collisions
via neutral H atoms $(S_{\rm H}=1)$, hereafter quoted as SH1.
For a single object (LP831-070)
only an upper limit to oxygen abundance has been determined.
Typical uncertanties are 
$\Delta[$Fe/H$]=\mp0.15$ and $\Delta[$O/H$]=\mp0.15$.
Standard deviations are not provided for each star.
For further details refer to the parent paper (Fal09).

\begin{figure*}[t]
\begin{center}      
\includegraphics[scale=0.8]{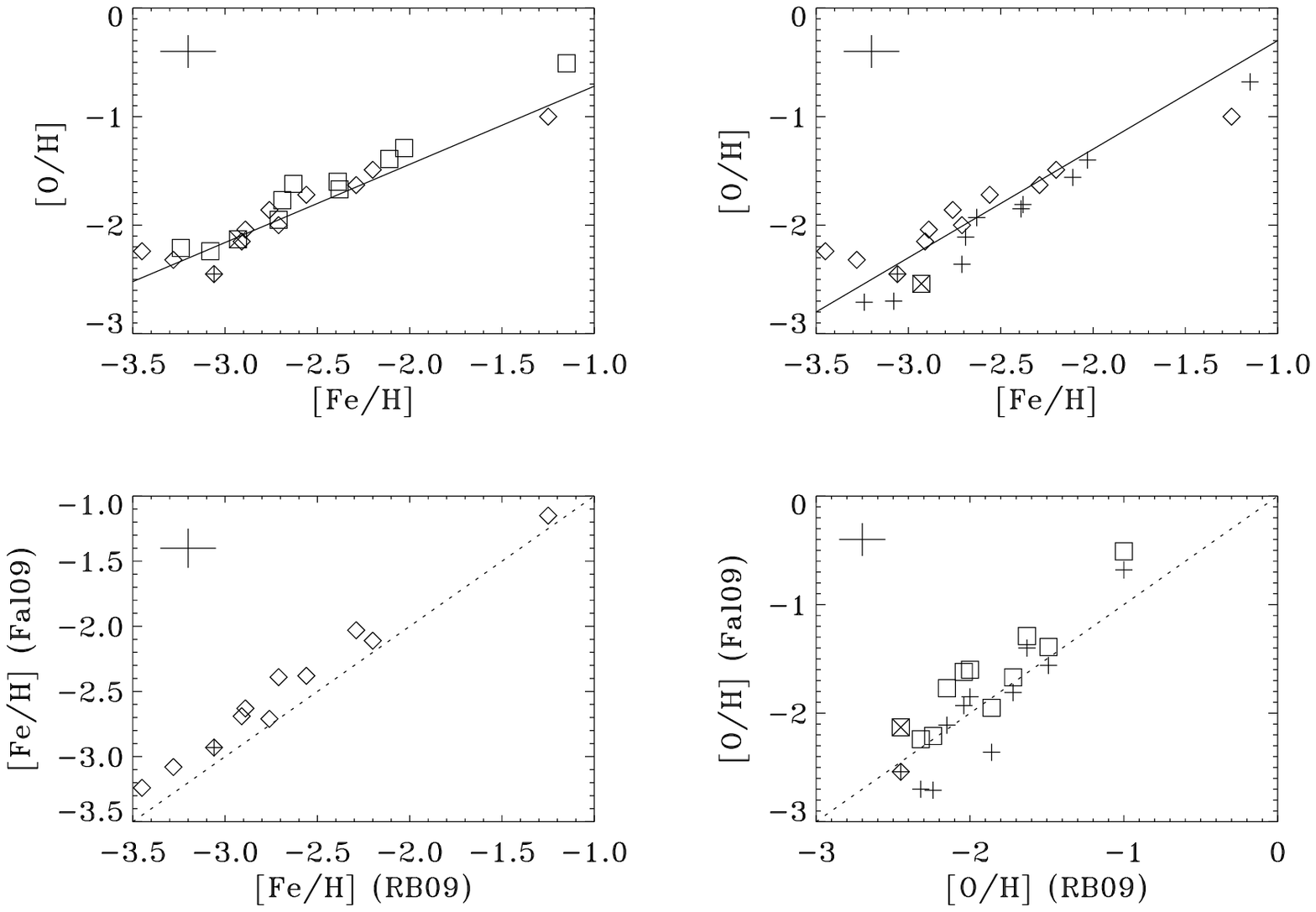}                      
\caption[rbfa]{Comparison between (i) [O/H]-[Fe/H] empirical
relations deduced from 11 halo stars in common among
RB09 (upper panels, diamonds) and Fal09, case LTE
(top left, squares), case SH1 (top right, crosses),
and (ii) [Fe/H] (bottom left) and [O/H] (bottom
right), case LTE (squares), case SH1 (crosses),
deduced from the above mentioned stars.
The composite symbols mark LP831-070, where only an
upper limit to oxygen abundance is available in the
Fal09 sample.   The straight lines in upper panels
are [O/H$]=0.72[$Fe/H] (left) and [O/H$]=[$Fe/H$]+
0.70$ (right).
The 1:1 correspondence in lower
panels is represented by a dashed line.   Typical
uncertainties are visualized as crosses on the top
left corner of each panel.}
\label{f:rbfa}
\end{center}       
\end{figure*}

The RB09 and Fal09 samples have $N=11$ (necessarily
halo) stars in common, where the values assumed for
effective temperature and surface gravity have been
determined using different methods, yielding different
values for each star, as shown in Table\,\ref{t:rbfa}.
\begin{table}
\caption{Values of effective temperature, $T_{\rm eff}$,
decimal logarithm of surface gravity, $\log g$, iron
abundance, [Fe/H], and oxygen abundance, [O/H], for $N=11$
stars in common between the RB09 and Fal09 sample.   With
regard to the latter, oxygen abundance has been determined
using three different methods, LTE, SH0, SH1, 
respectively,
and oxygen abundances related to LP831-070 are upper limits.
For further details refer to the text.}
\label{t:rbfa}
\begin{center}
\begin{tabular}{|l|l|l|l|l|l|l|l|l|l|l|} \hline
\multicolumn{1}{|c|}{} &
\multicolumn{4}{c|}{RN09} &
\multicolumn{6}{c|}{Fal09} \\
\multicolumn{1}{|c|}{star} &
\multicolumn{1}{c}{$T_{\rm eff}$} &
\multicolumn{1}{c}{$\log g$} &
\multicolumn{1}{c}{[Fe/H]} &
\multicolumn{1}{c|}{[O/H]} &
\multicolumn{1}{c}{$T_{\rm eff}$} &
\multicolumn{1}{c}{$\log g$} &
\multicolumn{1}{c}{[Fe/H]} &
\multicolumn{1}{c}{[O/H]} &
\multicolumn{1}{c}{[O/H]} &
\multicolumn{1}{c|}{[O/H]}  \\
\multicolumn{1}{|c|}{} &
\multicolumn{1}{c}{} &
\multicolumn{1}{c}{} &
\multicolumn{1}{c}{} &
\multicolumn{1}{c|}{LTE} &
\multicolumn{1}{c}{} &
\multicolumn{1}{c}{} &
\multicolumn{1}{c}{} &
\multicolumn{1}{c}{LTE} &
\multicolumn{1}{c}{SH0} &
\multicolumn{1}{c|}{SH1}  \\
\hline
LP651-04          & 6030 & 4.26 & $-$2.89 & $-$2.04 & 6371 & 4.20 & $-$2.63 & $-$1.62 & $-$2.21 & $-$1.93 \\
BD-13$^\circ$3442 & 6090 & 4.11 & $-$2.91 & $-$2.15 & 6366 & 3.99 & $-$2.69 & $-$1.77 & $-$2.39 & $-$2.11 \\
G11-44            & 5820 & 3.58 & $-$2.29 & $-$1.63 & 6178 & 4.35 & $-$2.03 & $-$1.29 & $-$1.47 & $-$1.40 \\
G64-12            & 6074 & 3.72 & $-$3.45 & $-$2.24 & 6435 & 4.26 & $-$3.24 & $-$2.21 & $-$3.10 & $-$2.71 \\
G64-37            & 6122 & 3.87 & $-$3.28 & $-$2.32 & 6432 & 4.24 & $-$3.08 & $-$2.24 & $-$3.12 & $-$2.70 \\
LP635-14          & 5932 & 3.57 & $-$2.71 & $-$2.00 & 6367 & 4.11 & $-$2.39 & $-$1.60 & $-$2.03 & $-$1.85 \\
LP815-43          & 6405 & 4.37 & $-$2.76 & $-$1.86 & 6483 & 4.21 & $-$2.71 & $-$1.95 & $-$2.70 & $-$2.36 \\
HD084937          & 6206 & 3.89 & $-$2.20 & $-$1.49 & 6357 & 4.07 & $-$2.11 & $-$1.39 & $-$1.64 & $-$1.56 \\
HD140283          & 5692 & 3.47 & $-$2.56 & $-$1.72 & 5849 & 3.72 & $-$2.38 & $-$1.67 & $-$1.91 & $-$1.81 \\
HD194598          & 5875 & 4.20 & $-$1.25 & $-$1.00 & 6020 & 4.30 & $-$1.15 & $-$0.51 & $-$0.75 & $-$0.68 \\
LP831-070         & 5985 & 4.75 & $-$3.06 & $-$2.45 & 6232 & 4.36 & $-$2.93 & $-$2.13 & $-$2.95 & $-$2.54 \\
\hline                            
\end{tabular}                     
\end{center}                      
\end{table}                       
A comparison between [O/H]-[Fe/H] empirical
relations deduced from the data listed in
Table \ref{t:rbfa}, in connection with
RB09 and Fal09 (case LTE, left; SH1, right)
samples, is presented in Fig.\,\ref{f:rbfa},
upper panels, while the correspondance
between [Fe/H] (left) and [O/H] (right),
deduced from the above mentioned data, is
shown in lower panels.
With regard to Fal09 sample, case SH0 has
not been considered, as it seems to overcorrect
LTE abundances and yield values of [O/Fe] which
are probably too low.   For further details
refer to the parent paper (Fal09).

It can be seen that the [O/H]-[Fe/H] empirical
relation is slightly affected by the parent
sample in the LTE case, with the exception of
the most metal-rich sample object (Fig.\,\ref
{f:rbfa}, top left panel).   A low discrepancy appears
in the SH1 case (Fig.\,\ref{f:rbfa}, top right panel),
where results in absence of LTE approximation
are available for the Fal09 sample only.  The
occurrence of a systematic difference in [Fe/H]
determinations is probably due to different
methods related to different samples, which
makes all points lie above the straight line with
unit slope (Fig.\,\ref{f:rbfa}, bottom left panel).
The same holds, though to a lesser extent, for
[O/H] determinations related to LTE case
(Fig.\,\ref{f:rbfa}, bottom right panel, squares),
while points related to the SH1 case
(Fig.\,\ref{f:rbfa}, bottom right panel, crosses)
lie both above and below the straight line
with unit slope.

\begin{figure*}[t]
\begin{center}      
\includegraphics[scale=0.8]{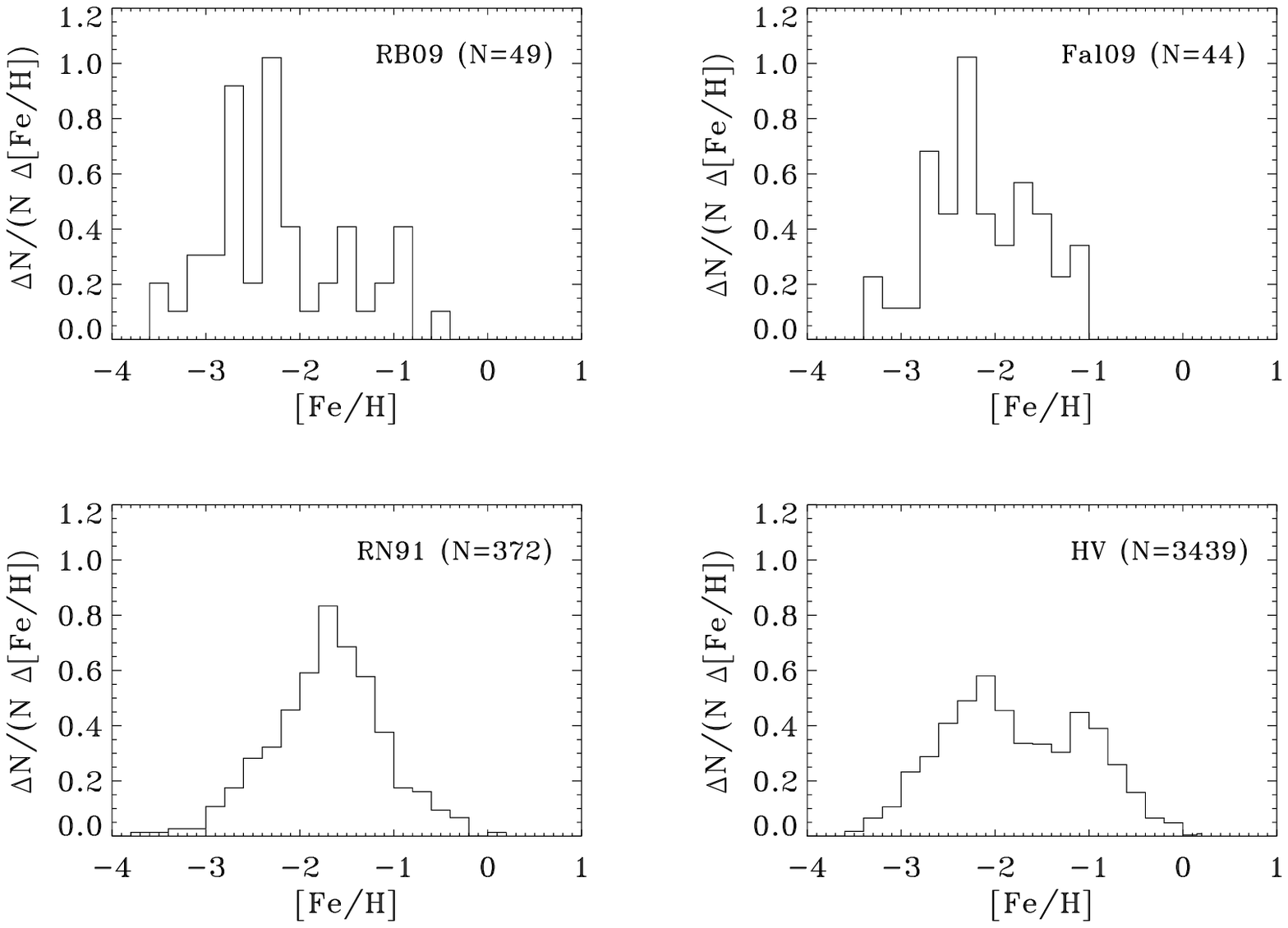}                      
\caption{[Fe/H] distributions (normalized to
the unit area) related to four different samples
discussed in the text. Sample denomination and
population are indicated on each panel.    In
all cases, the bin width is $\Delta[$Fe/H$]=0.2$.}
\label{f:FeH}
\end{center}       
\end{figure*}

The normalized [Fe/H] distributions related to the
RB09 and Fal09 sample are plotted in Fig.\,\ref
{f:FeH} (upper panels), where a similar trend
is shown and, in particular, both exhibit a
peak near [Fe/H$]=-2.2$, similar to their
counterpart deduced from the H\,V sample
(bottom right panel) but in contrast with their
counterpart deduced from the RN91 sample
(bottom left panel).
Then the RB09 and Fal09 samples seem to be
more representative of the outer halo than
the inner halo, where the most populated
metallicity bin is placed at [Fe/H$]\approx
-2.2$ and [Fe/H$]\approx-1.6$, respectively
(Carollo et al. 2007, 2010).   Accordingly, the
determination of the EDOD from the [Fe/H]
distribution related to RN91 and H\,V
samples has to be made under the assumption
that the [O/H]-[Fe/H] empirical relations,
deduced from the RB09 and Fal09 samples,
hold to a similar extent for both the inner
and the outer halo.

The following regression models are used
for fitting to [O/H]-[Fe/H] empirical relations.
\begin{description}
\item[G\hspace{4.4mm}]
heteroscedastic functional models where the
errors in X and in Y are uncorrelated (York
1966).
\item[Y\hspace{4.4mm}]
homoscedastic structural models where the
errors in X are negligible (ideally null)
with respect to the errors in Y (Isobe et
al. 1990, OLS(Y$\mid$X) therein%
\footnote{The captions, OLS(Y$\mid$X) and
OLS(X$\mid$Y),
have to be interchanged one with the other
in Table 1 of the quoted paper.}).
\item[X\hspace{4.4mm}]
homoscedastic structural models where the
errors in Y are negligible (ideally null)
with respect to the errors in X (Isobe et
al. 1990, OLS(X$\mid$Y) therein%
\footnote{The captions, OLS(X$\mid$Y) and
OLS(Y$\mid$X),
have to be interchanged one with the other
in Table 1 of the quoted paper.}).
\item[B\hspace{4.4mm}]
homoschedastic structural models where the
regression line bisects the angled formed
by the regression lines related to models
Y and X above (Isobe et al. 1990, OLS bisector
therein).
\item[O\hspace{4.4mm}]
homoscedastic structural or functional models
where the regression line minimizes the sum
of the perpendicular distances between the
data points and the line (Isobe et al. 1990;
Feigelson and Babu 1992; orthogonal regression
therein).
\item[R\hspace{4.4mm}]
homoscedastic structural models where the
regression line has slope equal to the
geometric mean of the slopes of the regression
lines related to models Y and X above (Isobe
et al. 1990, reduced major-axis therein).
\end{description}

\begin{figure*}[t]
\begin{center}      
\includegraphics[scale=0.8]{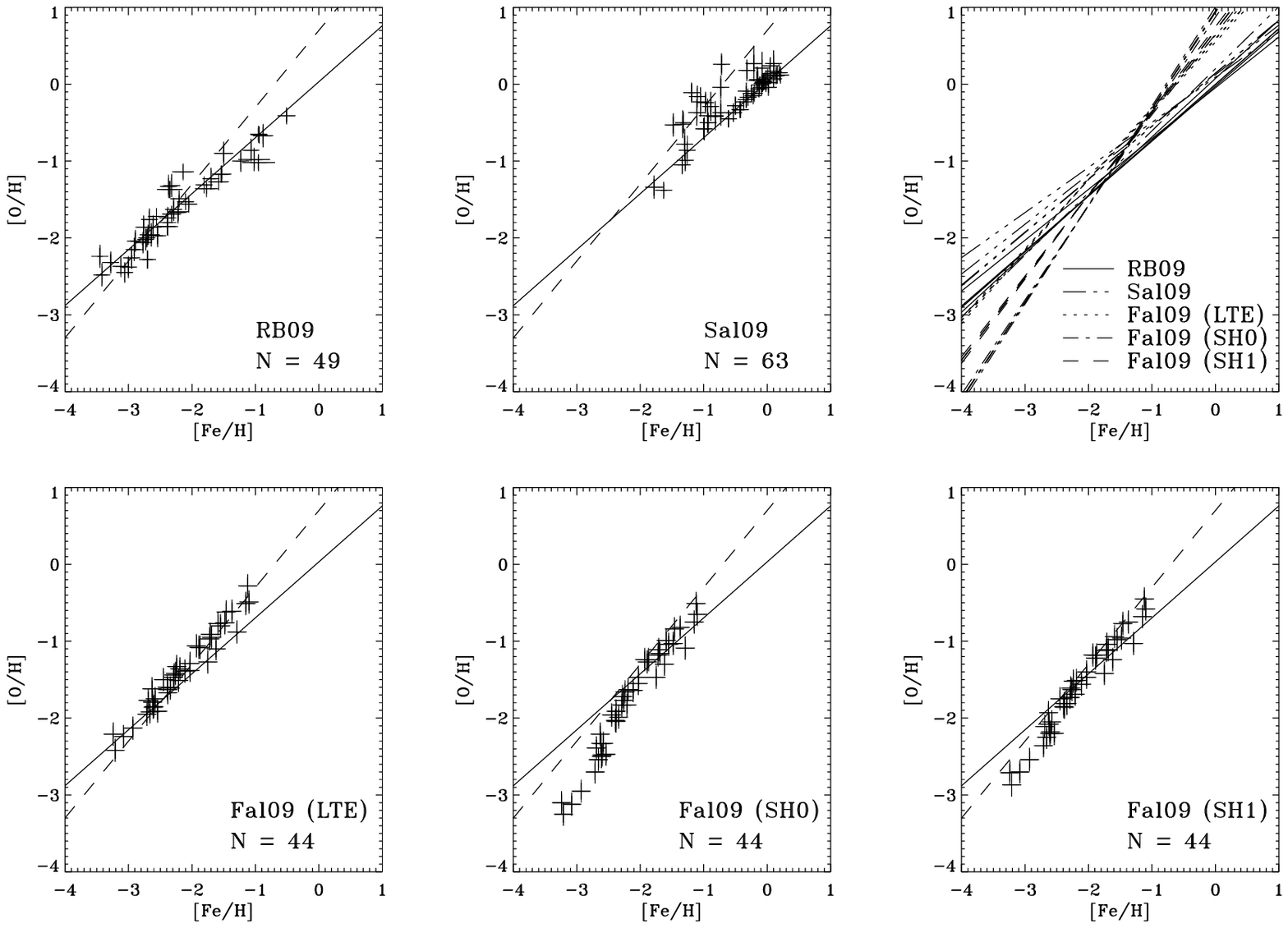}                      
\caption{The [O/H]-[Fe/H] empirical relation
deduced from two samples with heteroscedastic
data, RB09 and Sal09, and three samples with
homoscedastic data, Fal09, cases LTE, SH0,
and SH1, indicated on each panel together
with related population.   Also plotted on
each panel are the adopted linear dependences,
[O/H$]=0.72 [$Fe/H] (full) and [O/H$]=[$Fe/H$]+
0.70$ (dashed).   The regression lines
related to five different methods are shown for
each sample on the top right panel.   For further
details refer to the text.}
\label{f:FeHOH}
\end{center}       
\end{figure*}

The [O/H]-[Fe/H] empirical relations deduced
from RB09, Fal09 (cases LTE, SH0, SH1), and
Sal09 samples, shown in Fig.\,\ref{f:FeHOH},
are interpolated using the regression models
listed above.
For heteroscedastic data and homoscedastic
models, the typical uncertainties are
assigned to all the data points.   The
results are listed in Table \ref{t:inte}.
\begin{table}
\caption{Regression line slope and intercept
estimators, $\hat{a}$ and $\hat{b}$, and
related dispersion estimators, $\hat{\sigma}_
{\hat{a}}$, and $\hat{\sigma}_{\hat{b}}$,
for the models ($m$) mentioned in the text, applied
to the [O/H]-[Fe/H] empirical relation
deduced from the following samples: RB09
(top left); Fal09, cases LTE (top right),
SH0 (middle right), SH1 (bottom right);
and Sal09 (middle left).}
\label{t:inte}
\begin{center}
\begin{tabular}{lllllllll} \hline
\multicolumn{1}{c}{$m$} &
\multicolumn{1}{c}{$\hat{a}$} &
\multicolumn{1}{c}{$\hat{\sigma}_{\hat{a}}$} &
\multicolumn{1}{c}{$\hat{b}$} &
\multicolumn{1}{c}{$\hat{\sigma}_{\hat{b}}$} &
\multicolumn{1}{c}{$\hat{a}$} &
\multicolumn{1}{c}{$\hat{\sigma}_{\hat{a}}$} &
\multicolumn{1}{c}{$\hat{b}$} &
\multicolumn{1}{c}{$\hat{\sigma}_{\hat{b}}$}  \\
\hline

  &        &        &           &        &        &        &        &        \\
G & 0.7279 & 0.0294 & $+$0.0043 & 0.0672 & 0.9150 & 0.0305 & 0.5877 & 0.0666 \\
Y & 0.6917 & 0.0268 & $-$0.0766 & 0.0598 & 0.8961 & 0.0333 & 0.5476 & 0.0761 \\
X & 0.7600 & 0.0326 & $+$0.0742 & 0.0811 & 0.9381 & 0.0294 & 0.6366 & 0.0665 \\
B & 0.7253 & 0.0278 & $-$0.0025 & 0.0665 & 0.9169 & 0.0310 & 0.5916 & 0.0706 \\
O & 0.7143 & 0.0282 & $-$0.0268 & 0.0658 & 0.9150 & 0.0319 & 0.5877 & 0.0725 \\
R & 0.7251 & 0.0278 & $-$0.0030 & 0.0664 & 0.9168 & 0.0310 & 0.5916 & 0.0706 \\
  &        &        &           &        &        &        &        &        \\
G & 0.6383 & 0.0435 & $+$0.0619 & 0.0251 & 1.2640 & 0.0436 & 0.9519 & 0.0953 \\
Y & 0.5868 & 0.0596 & $+$0.0908 & 0.0211 & 1.2261 & 0.0459 & 0.8717 & 0.1017 \\
X & 0.8077 & 0.0563 & $+$0.2011 & 0.0422 & 1.2884 & 0.0434 & 1.0037 & 0.1003 \\
B & 0.6916 & 0.0513 & $+$0.1431 & 0.0282 & 1.2568 & 0.0441 & 0.9367 & 0.0998 \\
O & 0.6476 & 0.0620 & $+$0.1212 & 0.0268 & 1.2640 & 0.0449 & 0.9519 & 0.1019 \\
R & 0.6885 & 0.0523 & $+$0.1416 & 0.0279 & 1.2569 & 0.0441 & 0.9369 & 0.0998 \\
  &        &        &           &        &        &        &        &        \\
G &        &        &           &        & 1.0732 & 0.0343 & 0.7027 & 0.0750 \\
Y &        &        &           &        & 1.0492 & 0.0358 & 0.6518 & 0.0808 \\
X &        &        &           &        & 1.0946 & 0.0315 & 0.7479 & 0.0730 \\
B &        &        &           &        & 1.0716 & 0.0332 & 0.6993 & 0.0760 \\
O &        &        &           &        & 1.0732 & 0.0337 & 0.7027 & 0.0772 \\
R &        &        &           &        & 1.0716 & 0.0332 & 0.6993 & 0.0760 \\
\hline                            
\end{tabular}                     
\end{center}                      
\end{table}                       
An inspection of Fig.\,\ref{f:FeHOH} and 
Table 2 shows that systematic errors
related to different methods and/or
different spectral lines in oxygen
abundance determination, are dominant 
on both the dispersion due to measurement
errors and intrinsic scatter.   At this
stage, a precise fit to the data extracted
from a selected sample would be meaningless,
and only acceptable fits related to
different situations shall be considered.
Accordingly, the following
[O/H]-[Fe/H] empirical relations:
\begin{equation}
\label{eq:OF72}
{\rm [O/H]}=0.72{\rm [Fe/H]}~~;
\end{equation}
with regard to the RB09 sample,
Fig.\,\ref{f:FeHOH} (full lines
on data points), and
\begin{equation}
\label{eq:OF70}
{\rm [O/H]}={\rm [Fe/H]}+0.70~~;
\end{equation}
with regard to the Fal09 sample, case SH1, 
Fig.\,\ref{f:FeHOH} (dashed lines on data points),
shall be used for determining
the EDOD from the [Fe/H] distribution,
related to a selected sample.
The regression lines
related to five different methods listed
in Table \ref{t:inte}, are shown for
each sample on the top right panel of 
Fig.\,\ref{f:FeHOH}.

\subsection{The empirical differential oxygen abundance
distribution (EDOD)} \label{EDOD}

With regard to a selected [O/H]-[Fe/H] 
empirical relation:
\begin{equation}
\label{eq:OFg}
{\rm [O/H]}=a{\rm [Fe/H]}+b~~;
\end{equation}
and a specified [Fe/H] distribution, let
[Fe/H], [Fe/H]$^\mp$, be the coordinates
of a selected bin centre and bin left ($-$)
and right (+) extremum, respectively, and
[O/H], [O/H]$^\mp$ their counterparts 
deduced via Eq.\,(\ref{eq:OFg}) for the
related [O/H] distribution.   Accordingly,
the bin semiamplitude reads:
\begin{equation}
\label{eq:DpmO}
\Delta^{\mp}{\rm [O/H]}=\frac{{\rm[O/H]}^+-{\rm[O/H]}^-}2=
a\frac{{\rm[Fe/H]}^+-{\rm[Fe/H]}^-}2=a\Delta^{\mp}{\rm [Fe/H]}~~;
\end{equation}
where [Fe/H] and [O/H] are logarithmic number abundances
normalized to the solar value (e.g., C07).

The oxygen mass abundance normalized to the
solar value, $\phi$, to a good extent may
be expressed as (e.g., C07):
\begin{equation}
\label{eq:phi}
\phi=\frac{Z_{\rm O}}{(Z_{\rm O})_\odot}=\exp_{10}{\rm [O/H]}~~;
\end{equation}
and a selected bin centre and bin
semiamplitude, related to the [O/H]
distribution, read (e.g., C07):
\begin{lefteqnarray}
\label{eq:phic}
&& \phi=\frac12\{\exp_{10}{\rm[O/H]}^++\exp_{10}{\rm[O/H]}^-\}~~; \\
\label{eq:phia}
&& \Delta^\mp\phi=\frac12\{\exp_{10}{\rm[O/H]}^+-\exp_{10}{\rm[O/H]}^-\}~~;
\end{lefteqnarray}
where the bin width is variable for a
constant bin width related to the [O/H]
distribution.   The [O/H]-[Fe/H] and
$\phi$-[Fe/H] empirical relations,
expressed by Eqs.\,(\ref{eq:OF72}) and
(\ref{eq:OF70}), by use of
Eqs.\,(\ref{eq:phic}) and
(\ref{eq:phia}) in the latter case,
are represented in Table \ref{t:FeHOHp},
where $B_{\rm F}=[$Fe/H] and $B_{\rm O}=[$O/H]
to save space.
\begin{table}
\caption[par]{The normalized oxygen abundance,
$\phi=Z_{\rm O}/(Z_{\rm O})_\odot$,
for two different [O/H]-[Fe/H] empirical relations
deduced from interpolation
to two different data sets, namely
RB09 in presence of the local thermodynamic
equilibrium approximation (left side), and Fal09 in
absence of the local thermodynamic equilibrium
approximation, case SH1 (right side).
Upper and lower values for
each bin are denoted by the apex + and $-$, respectively.
The corresponding bin mean value and semiamplitude,
with regard to normalized oxygen abundance, are
labelled as $\phi$ and $\Delta^\mp\phi$, respectively.
Labels $B_{\rm F}^\mp$ and $B_{\rm O}^\mp$ stand for [Fe/H]$
^\mp$ and [O/H]$^\mp$, respectively.}
\label{t:FeHOHp}
\begin{center}
\begin{tabular}{llllllllll}
\multicolumn{2}{c|}{}
& \multicolumn{4}{c|}{[O/H] = 0.72 [Fe/H]\hspace{10mm}(RB09)}
& \multicolumn{4}{c}{[O/H] = [Fe/H] + 0.70\hspace{10mm}(Fal09)} \\
\hline\noalign{\smallskip}
\multicolumn{1}{c}{$B_{\rm F}^-$} &
\multicolumn{1}{c}{$B_{\rm F}^+$} &
\multicolumn{1}{c}{$B_{\rm O}^-$} &
\multicolumn{1}{c}{$B_{\rm O}^+$} &
\multicolumn{1}{c}{$\phi$} &
\multicolumn{1}{c}{$\Delta^\mp\phi$} &
\multicolumn{1}{c}{$B_{\rm O}^-$} &
\multicolumn{1}{c}{$B_{\rm O}^+$} &
\multicolumn{1}{c}{$\phi$} &
\multicolumn{1}{c}{$\Delta^\mp\phi$} \\
\noalign{\smallskip}
\hline\noalign{\smallskip}
$-$4.2 & $-$4.0 & $-$3.024 & $-$2.880 & 1.1322D$-$3 & 1.8601D$-$4 & $-$3.5 & $-$3.3 & 4.0871D$-$4 & 9.2480D$-$5 \\
$-$4.0 & $-$3.8 & $-$2.880 & $-$2.736 & 1.5774D$-$3 & 2.5914D$-$4 & $-$3.3 & $-$3.1 & 6.4776D$-$4 & 1.4657D$-$4 \\
$-$3.8 & $-$3.6 & $-$2.736 & $-$2.592 & 2.1976D$-$3 & 3.6102D$-$4 & $-$3.1 & $-$2.9 & 1.0266D$-$3 & 2.3230D$-$4 \\
$-$3.6 & $-$3.4 & $-$2.592 & $-$2.448 & 3.0615D$-$3 & 5.0296D$-$4 & $-$2.9 & $-$2.7 & 1.6271D$-$3 & 3.6817D$-$4 \\
$-$3.4 & $-$3.2 & $-$2.448 & $-$2.304 & 4.2652D$-$3 & 7.0071D$-$4 & $-$2.7 & $-$2.5 & 2.5788D$-$3 & 5.8351D$-$4 \\
$-$3.2 & $-$3.0 & $-$2.304 & $-$2.160 & 5.9421D$-$3 & 9.7619D$-$4 & $-$2.5 & $-$2.3 & 4.0871D$-$3 & 9.2480D$-$4 \\
$-$3.0 & $-$2.8 & $-$2.160 & $-$2.016 & 8.2783D$-$3 & 1.3600D$-$3 & $-$2.3 & $-$2.1 & 6.4776D$-$3 & 1.4657D$-$3 \\
$-$2.8 & $-$2.6 & $-$2.016 & $-$1.872 & 1.1533D$-$2 & 1.8947D$-$3 & $-$2.1 & $-$1.9 & 1.0266D$-$2 & 2.3230D$-$3 \\
$-$2.6 & $-$2.4 & $-$1.872 & $-$1.728 & 1.6067D$-$2 & 2.6396D$-$3 & $-$1.9 & $-$1.7 & 1.6271D$-$2 & 3.6817D$-$3 \\
$-$2.4 & $-$2.2 & $-$1.728 & $-$1.584 & 2.2384D$-$2 & 3.6774D$-$3 & $-$1.7 & $-$1.5 & 2.5788D$-$2 & 5.8351D$-$3 \\
$-$2.2 & $-$2.0 & $-$1.584 & $-$1.440 & 3.1185D$-$2 & 5.1231D$-$3 & $-$1.5 & $-$1.3 & 4.0871D$-$2 & 9.2480D$-$3 \\
$-$2.0 & $-$1.8 & $-$1.440 & $-$1.296 & 4.3445D$-$2 & 7.1373D$-$3 & $-$1.3 & $-$1.1 & 6.4776D$-$2 & 1.4657D$-$2 \\
$-$1.8 & $-$1.6 & $-$1.296 & $-$1.152 & 6.0526D$-$2 & 9.9434D$-$3 & $-$1.1 & $-$0.9 & 1.0266D$-$1 & 2.3230D$-$2 \\
$-$1.6 & $-$1.4 & $-$1.152 & $-$1.008 & 8.4322D$-$2 & 1.3853D$-$2 & $-$0.9 & $-$0.7 & 1.6271D$-$1 & 3.6817D$-$2 \\
$-$1.4 & $-$1.2 & $-$1.008 & $-$0.864 & 1.1747D$-$1 & 1.9299D$-$2 & $-$0.7 & $-$0.5 & 2.5788D$-$1 & 5.8351D$-$2 \\
$-$1.2 & $-$1.0 & $-$0.864 & $-$0.720 & 1.6366D$-$1 & 2.6887D$-$2 & $-$0.5 & $-$0.3 & 4.0871D$-$1 & 9.2480D$-$2 \\
$-$1.0 & $-$0.8 & $-$0.720 & $-$0.576 & 2.2800D$-$1 & 3.7457D$-$2 & $-$0.3 & $-$0.1 & 6.4776D$-$1 & 1.4657D$-$1 \\
$-$0.8 & $-$0.6 & $-$0.576 & $-$0.432 & 3.1764D$-$1 & 5.2184D$-$2 & $-$0.1 & $+$0.1 & 1.0266D$+$0 & 2.3230D$-$1 \\
$-$0.6 & $-$0.4 & $-$0.432 & $-$0.288 & 4.4253D$-$1 & 7.2700D$-$2 & $+$0.1 & $+$0.3 & 1.6271D$+$0 & 3.6817D$-$1 \\
$-$0.4 & $-$0.2 & $-$0.288 & $-$0.144 & 6.1651D$-$1 & 1.0128D$-$1 & $+$0.3 & $+$0.5 & 2.5788D$+$0 & 5.8351D$-$1 \\
$-$0.2 & $+$0.0 & $-$0.144 & $+$0.000 & 8.5890D$-$1 & 1.4110D$-$1 & $+$0.5 & $+$0.7 & 4.0871D$+$0 & 9.2480D$-$1 \\
$+$0.0 & $+$0.2 & $+$0.000 & $+$0.144 & 1.1966D$-$0 & 1.9658D$-$1 & $+$0.7 & $+$0.9 & 6.4776D$+$0 & 1.4657D$-$0 \\
$+$0.2 & $+$0.4 & $+$0.144 & $+$0.288 & 1.6670D$-$0 & 2.7386D$-$1 & $+$0.9 & $+$1.1 & 1.0266D$+$1 & 2.3230D$-$0 \\
\noalign{\smallskip}
\hline              
\end{tabular}       
\end{center}        
\end{table}         

Following recent attempts (Rocha-Pinto and
Maciel 1996; C00; C01; C07), the EDOD in a
selected class of objects is defined as:
\begin{equation}
\label{eq:psie}
\psi(\phi)=\log\frac{\Delta N}{N\Delta\phi}~~;
\end{equation}
where $\Delta N$ is the number of objects
within a normalized oxygen abundance bin,
$\Delta\phi$, centered in $\phi$, and $N$
is the number of sample objects.   The
increment ratio, $\Delta N/\Delta\phi$,
used in earlier attempts (Pagel 1989;
Malinie et al. 1993) is replaced
by its normalized counterpart, $\Delta N/
(N\Delta\phi)$, to allow comparison between
different samples.   The uncertainty on
$\Delta N$ is evaluated from Poisson errors
(e.g., RN91), as $\sigma_{\Delta N}=(\Delta
N)^{1/2}$, and the related uncertainty in
the EDOD is:
\begin{leftsubeqnarray}
\slabel{eq:Dpsiea}
&& \Delta^\mp\psi=\vert\psi-\psi^\mp\vert=
\log\left[1\mp\frac{(\Delta N)^
{1/2}}{\Delta N}\right]~~; \\
\slabel{eq:Dpsieb}
&& \psi^\mp=\log\frac{\Delta N\mp(\Delta N)^{1/2}}{N\Delta\phi}~~;
\label{seq:Dpsie}
\end{leftsubeqnarray}
where $\psi^-\rightarrow-\infty$ in the limit
$\Delta N\rightarrow1$.   For further details
refer to the parent papers (C01; C07).

\begin{figure*}[t]  
\begin{center}      
\includegraphics[scale=0.8]{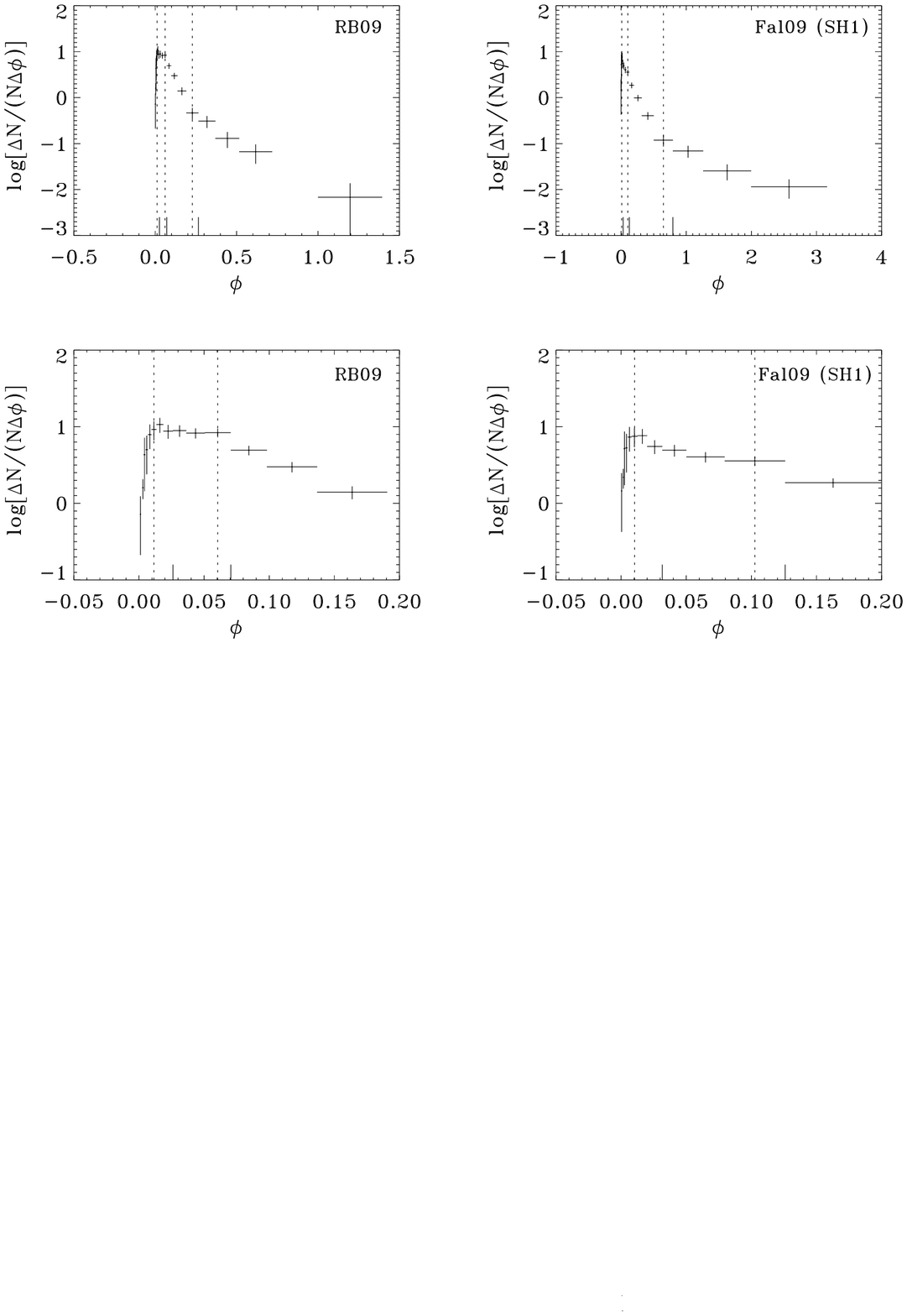}                      
\caption[ddbb]{The empirical, differential oxygen abundance
distribution (EDOD) related to the fs10 sample for
[O/H]-[Fe/H] empirical relations deduced from the RB09
(left panels) and Fal09, case SH1 (right panels) sample.
The whole distribution is represented in upper panels,
while lower panels zoom the low-metallicity
range.   The uncertainty of the distribution is determined
from Poisson errors.   The vertical bars on the horisontal
axis mark [Fe/H$]=-2.2$, $-$1.6, and $-$0.8.   The vertical
dotted lines mark [Fe/H$]=-2.7$, $-$1.7, and $-$0.9, where
the linear trend of the EDOD changes passing from a region
to the adjacent one.   For further details refer to the text.}
\label{f:ddbb}     
\end{center}       
\end{figure*}

The EDOD related to RN91, H\,V, and fs10 samples
are listed in Table \ref{t:psph} for the
[O/H]-[Fe/H] empirical relations expressed
by Eqs.\,(\ref{eq:OF72}) and (\ref{eq:OF70}),
left and right side, respectively.
\begin{table}
\caption{The empirical, differential
oxygen abundance distribution (EDOD) in the
inner halo,
deduced from the fs10 sample
($N=7452$) using two different
[O/H]-[Fe/H] empirical relations, determined from
interpolation to two different data sets,
RB09 in presence of the local thermodynamic
equilibrium approximation (left side), and Fal09 in
absence of the local thermodynamic equilibrium
approximation, case SH1 (right side).
The fictitious fs10 sample results from the
combination of the H\,V sample ($N=3439$) for
lower metallicities, $-4.2\le$[Fe/H]$<-3.0$, 
and the RN91 sample ($N=372$) for higher metallicities,
$-2.8<$[Fe/H]$\le+0.2$, under the assumption that
the two samples are equally representative of the
inner halo within the metallicity bin, $-3.0\le$
[Fe/H]$\le-2.8$. The error on the generic bin height
has been estimated from the Poissonian error of its
counterpart related to the parent sample.   See text
for further details.}
\label{t:psph}
\begin{center}
\begin{tabular}{llllllllrrr} \hline
\multicolumn{4}{c|}{[O/H] = 0.72 [Fe/H]\hspace{10mm}(RB09)}
&\multicolumn{4}{c|}{[O/H] = [Fe/H] + 0.70\hspace{10mm}(Fal09)}
&\multicolumn{3}{c}{$\Delta N$} \\
\hline\noalign{\smallskip}
\multicolumn{1}{c}{$\phi$} & \multicolumn{1}{c}{$\phantom{0}\psi$} &
\multicolumn{1}{c}{$\Delta^-\psi$} & \multicolumn{1}{c}{$\Delta^+\psi$} &
\multicolumn{1}{c}{$\phi$} & 
\multicolumn{1}{c}{$\phantom{0}\psi$} &
\multicolumn{1}{c}{$\Delta^-\psi$} & \multicolumn{1}{c}{$\Delta^+\psi$} &
\multicolumn{1}{c}{fs10}  & 
\multicolumn{1}{c}{H\,V}  & 
\multicolumn{1}{c}{RN91}  \\
\noalign{\smallskip}
\hline\noalign{\smallskip}                                                                                           
1.1322D$-$3 & $-$1.4181D$-$1 & 5.3329D$-$1 & 2.3226D$-$1 & 4.0871D$-$4 & $+$1.6168D$-$1 & 5.3329D$-$1 & 2.3226D$-$1 &    2 &   2 &  0 \\
1.5774D$-$3 &                &             &             & 6.4776D$-$4 &                &             &             &    0 &   0 &  0 \\
2.1976D$-$3 &                &             &             & 1.0266D$-$3 &                &             &             &    0 &   0 &  1 \\
3.0615D$-$3 & $+$2.0434D$-$1 & 1.4793D$-$1 & 1.1014D$-$1 & 1.6271D$-$3 & $+$3.3983D$-$1 & 1.4793D$-$1 & 1.1014D$-$1 &   12 &  12 &  1 \\
4.2652D$-$3 & $+$6.3437D$-$1 & 4.7712D$-$1 & 2.2185D$-$1 & 2.5788D$-$3 & $+$7.1386D$-$1 & 4.7712D$-$1 & 2.2185D$-$1 &   45 &  45 &  2 \\
5.9421D$-$3 & $+$7.0048D$-$1 & 3.2187D$-$1 & 1.8282D$-$1 & 4.0871D$-$3 & $+$7.2397D$-$1 & 3.2187D$-$1 & 1.8282D$-$1 &   73 &  73 &  2 \\
8.2783D$-$3 & $+$8.9728D$-$1 & 1.8947D$-$1 & 1.3148D$-$1 & 6.4776D$-$3 & $+$8.6477D$-$1 & 1.8947D$-$1 & 1.3148D$-$1 &  160 & 160 &  8 \\
1.1533D$-$2 & $+$9.6413D$-$1 & 1.4107D$-$1 & 1.0631D$-$1 & 1.0266D$-$2 & $+$8.7562D$-$1 & 1.4107D$-$1 & 1.0631D$-$1 &  260 & 198 & 13 \\
1.6067D$-$2 & $+$1.0284D$-$0 & 1.0691D$-$1 & 8.5725D$-$2 & 1.6271D$-$2 & $+$8.8390D$-$1 & 1.0691D$-$1 & 8.5725D$-$2 &  420 & 281 & 21 \\
2.2384D$-$2 & $+$9.4240D$-$1 & 9.9155D$-$2 & 8.0671D$-$2 & 2.5788D$-$2 & $+$7.4189D$-$1 & 9.9155D$-$2 & 8.0671D$-$2 &  480 & 337 & 24 \\
3.1185D$-$2 & $+$9.4967D$-$1 & 8.1707D$-$2 & 6.8742D$-$2 & 4.0871D$-$2 & $+$6.9316D$-$1 & 8.1707D$-$2 & 6.8742D$-$2 &  680 & 399 & 34 \\
4.3445D$-$2 & $+$9.1764D$-$1 & 7.0967D$-$2 & 6.0983D$-$2 & 6.4776D$-$2 & $+$6.0513D$-$1 & 7.0967D$-$2 & 6.0983D$-$2 &  880 & 313 & 44 \\
6.0526D$-$2 & $+$9.2258D$-$1 & 5.8986D$-$2 & 5.1924D$-$2 & 1.0266D$-$1 & $+$5.5407D$-$1 & 5.8986D$-$2 & 5.1924D$-$2 & 1240 & 231 & 62 \\
8.4322D$-$2 & $+$6.9376D$-$1 & 6.5516D$-$2 & 5.6916D$-$2 & 1.6271D$-$1 & $+$2.6925D$-$1 & 6.5516D$-$2 & 5.6916D$-$2 & 1020 & 229 & 51 \\
1.1747D$-$1 & $+$4.7566D$-$1 & 7.1860D$-$2 & 6.1640D$-$2 & 2.5788D$-$1 & $-$4.8510D$-$3 & 7.1860D$-$2 & 6.1640D$-$2 &  860 & 209 & 43 \\
1.6366D$-$1 & $+$1.4535D$-$1 & 9.0970D$-$2 & 7.5175D$-$2 & 4.0871D$-$1 & $-$3.9116D$-$1 & 9.0970D$-$2 & 7.5175D$-$2 &  560 & 308 & 28 \\
2.2800D$-$1 & $-$3.3187D$-$1 & 1.4107D$-$1 & 1.0631D$-$1 & 6.4776D$-$1 & $-$9.2438D$-$1 & 1.4107D$-$1 & 1.0631D$-$1 &  260 & 268 & 13 \\
3.1764D$-$1 & $-$5.1063D$-$1 & 1.4793D$-$1 & 1.1014D$-$1 & 1.0266D$+$0 & $-$1.1591D$-$0 & 1.4793D$-$1 & 1.1014D$-$1 &  240 & 178 & 12 \\
4.4253D$-$1 & $-$8.8871D$-$1 & 2.0618D$-$1 & 1.3924D$-$1 & 1.6271D$+$0 & $-$1.5932D$-$0 & 2.0618D$-$1 & 1.3924D$-$1 &  140 & 109 &  7 \\
6.1651D$-$1 & $-$1.1788D$-$0 & 2.5744D$-$1 & 1.6053D$-$1 & 2.5788D$+$0 & $-$1.9393D$-$0 & 2.5744D$-$1 & 1.6053D$-$1 &  100 &  45 &  5 \\
8.5890D$-$1 &                &             &             & 4.0871D$+$0 &                &             &             &    0 &  33 &  0 \\
1.1966D$-$0 & $-$2.1658D$+$0 & $+\infty$   & 3.0103D$-$1 & 6.4776D$+$0 & $-$3.0383D$-$0 & $+\infty$   & 3.0103D$-$1 &   20 &   3 &  1 \\
1.6670D$-$0 &                &             &             & 1.0266D$+$1 &                &             &             &    0 &   6 &  0 \\
\noalign{\smallskip}
\hline                                                      
\end{tabular}                                               
\end{center}                                                
\end{table}                                                 
The EDOD related to the fs10 sample,
taken to be representative of the
inner halo, is plotted in Fig.\,\ref{f:ddbb}
for the [O/H]-[Fe/H] empirical relations expressed
by Eqs.\,(\ref{eq:OF72}) and (\ref{eq:OF70}),
left and right panel, respectively.
Upper panels represent the whole distribution, while
lower panels zoom the low-metallicity range.
The vertical bars on the horisontal
axis mark [Fe/H$]=-2.2$, $-$1.6, where the global
[Fe/H] distribution of the outer and inner halo,
respectively, peak according to recent results
(Carollo et al. 2007, 2010), and [Fe/H$]=-$0.8, where a
transition from halo to bulge/disk globular clusters
occurs (Mackey and van den Bergh 2005).

A main feature of the EDODs plotted in
Fig.\,\ref{f:ddbb}, is the presence
of five regions characterized by a
nearly linear trend, which shall be
named O, A, F, C, E, respectively,
and defined in the following metallicity
ranges, each containing $n_{\rm X}$ bins,
X=O, A, F, C, E.
\begin{description}
\item[O\hspace{4.4mm}]
[Fe/H$]<-4.2$; [Fe/H$]>-0.2$; $n_{\rm O}\to+
\infty$; the distribution coincides with
the horisontal axis after removing a single
high-metallicity star from the RN91 sample,
considered as due to disk contamination or,
in any case, an outlier.   On the other
hand, the last appears in related tables
and figures for comparison.
\item[A\hspace{4.4mm}]
$-4.2\le[$Fe/H]$\le-2.7$; $n_{\rm A}=6;$
the distribution is steep with positive
slope.
\item[F\hspace{4.4mm}]
$-2.7\le[$Fe/H$]\le-1.7$; $n_{\rm F}=6;$
the distribution is mild with negative
slope.
\item[C\hspace{4.4mm}]
$-1.7\le[$Fe/H]$\le-0.9$; $n_{\rm C}=5;$
the distribution is steep with negative
slope.
\item[E\hspace{4.4mm}]
$-0.9\le[$Fe/H$]\le-0.2$; $n_{\rm E}=4;$
the distribution is less steep with negative
slope.
\end{description}
In absence of LTE approximation with
regard to the [O/H]-[Fe/H] empirical
relation deduced from the Fal09 sample
(case SH1), regions F and C may merge
into a single region, FC, within the
range, $-2.7\le[$Fe/H$]\le-0.9$, 
including $n_{\rm FC}=10$ bins.
The vertical dashed lines in 
Fig.\,\ref{f:ddbb} mark [Fe/H$]=
-2.7$, $-$1.7, $-$0.8, from the
left to the right.   It can be
seen that the global [Fe/H]
distribution peaks early within
the F and C region for the outer
and inner halo, respectively.
In addition, data points on the
boundary between adjacent regions
follow the linear trend exhibited
by every of them.

The regression line related to the
EDOD within each populated region,
has been determined using the B
model listed in Table \ref{t:inte},
under the assumption that the
intrinsic scatter is dominant
(Isobe et al. 1990).   A single
high-metallicity star from the RN91 sample,
considered as due to disk contamination or,
in any case, an outlier, has not been included
in the fitting procedure.   The regression line
slope and intercept estimators and related
dispersion estimators are listed in Table
\ref{t:abs} for each region of the EDODs
plotted in Fig.\,\ref{f:ddbb}.   The results
are consistent (within $\mp\sigma$) with
their counterparts determined using the other
models listed in Table \ref{t:inte}.
\begin{table}
\caption{Regression line slope and intercept
estimators, $\hat{a}_{\rm B}$ and $\hat{b}_{\rm B}$, and
related dispersion estimators, $\hat{\sigma}_
{\hat{a}_{\rm B}}$, and $\hat{\sigma}_{\hat{b}_{\rm B}}$,
for the B model applied to the 
oxygen abundance distribution (EDOD)
plotted in Fig.\,\ref{f:ddbb} with regard to
different [O/H]-[Fe/H] empirical relations,
deduced from the RB09 sample in presence of LTE
approximation via Eq.\,(\ref{eq:OF72}) (top panel), 
and from the Fal09
sample in absence of LTE approximation, case SH1,
via Eq.\,(\ref{eq:OF70})
(bottom panel).   The method has been applied to
each region (X) separately.   Data points on the
boundary between adjacent regions are used
for determining regression lines within both
of them.}
\label{t:abs}
\begin{center}
\begin{tabular}{lllll} \hline
\multicolumn{1}{l}{X} &
\multicolumn{1}{c}{$\hat{a}_{\rm B}$} &
\multicolumn{1}{c}{$\hat{\sigma}_{\hat{a}_{\rm B}}$} &
\multicolumn{1}{c}{$\hat{b}_{\rm B}$} &
\multicolumn{1}{c}{$\hat{\sigma}_{\hat{b}_{\rm B}}$} \\
\hline
   &                 &              &                 &              \\
A  & $+$1.1382 E$+$2 & 2.0648 E$+$1 & $-$1.0567 E$-$1 & 1.2296 E$-$1 \\ 
F  & $-$2.0950 E$+$0 & 7.2725 E$-$1 & $+$1.0188 E$+$0 & 2.9109 E$-$2 \\ 
C  & $-$7.3565 E$+$0 & 1.2375 E$-$1 & $+$1.3433 E$+$0 & 2.3577 E$-$2 \\ 
E  & $-$2.2569 E$+$0 & 1.0098 E$-$1 & $+$1.7784 E$-$1 & 3.2148 E$-$2 \\ 
   &                 &              &                 &              \\
A  & $+$7.9432 E$+$1 & 1.9364 E$+$1 & $+$2.7571 E$-$1 & 9.2608 E$-$2 \\
F  & $-$3.8855 E$+$0 & 6.1570 E$-$1 & $+$8.9453 E$-$1 & 2.9691 E$-$2 \\
C  & $-$2.6460 E$+$0 & 1.4542 E$-$1 & $+$7.3661 E$-$1 & 5.9411 E$-$2 \\
FC & $-$2.8643 E$+$0 & 1.3796 E$-$1 & $+$8.2804 E$-$1 & 2.7988 E$-$2 \\                                     
E  & $-$5.3661 E$-$1 & 3.5681 E$-$2 & $-$6.1514 E$-$1 & 4.3965 E$-$2 \\
\hline       
\end{tabular}
\end{center} 
\end{table}  
The selection of a special method is of
little relevance, due to the paucity of
data within each region.

\begin{figure*}[t]  
\begin{center}      
\includegraphics[scale=0.8]{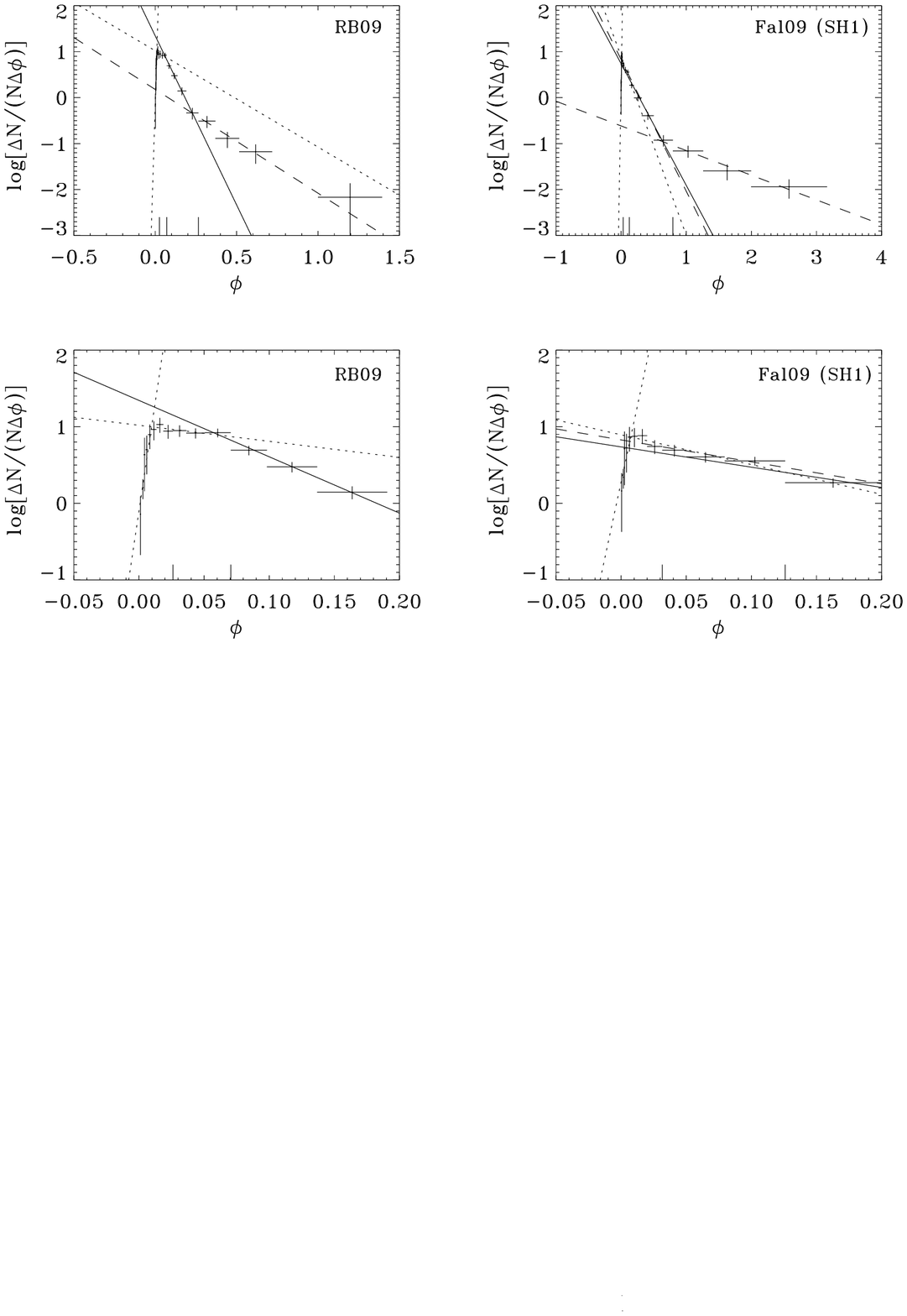}                      
\caption[ddbb]{Regression lines to the empirical
differential oxygen abundance distribution (EDOD) plotted
in Fig.\,\ref{f:ddbb}, with regard to the regions
(from the left to the right): A (dotted, positive
slope, $n_{\rm A}=6$), F(dotted, negative slope,
$n_{\rm F}=6$), C (full, $n_{\rm C}=5$),
E (dashed, $n_{\rm E}=4$).   The more inclined
dashed line on right panels corresponds to the
FC ($n_{\rm FC}=10$) region, in alternative to
F and C regions separately.   Other captions as
in Fig.\,\ref{f:ddbb}.   For further details
refer to the text.}
\label{f:tdbb}     
\end{center}       
\end{figure*}

The regression lines are represented in
Fig.\,\ref{f:tdbb} for each region (from
the left to the right): A (dotted, positive
slope), F(dotted, negative slope), C (full),
E (dashed), with regard to the EDOD plotted
in Fig.\,\ref{f:ddbb}.
In absence of LTE approximation, case
SH1 (right panels), a more inclined
dashed line fits to the FC region, in
alternative to F and C regions separately.
To ensure continuity, the above mentioned
regions must be redefined by the intersections
of regression lines, which make the transition
from a selected region to the adjacent one.
The results are listed in Table \ref{t:figs}
for the EDOD and related regression lines
plotted in Fig.\,\ref{f:tdbb}.
\begin{table}
\caption{Transition points between adjacent
regions, as determined from the intersection
of related regression lines, for the
oxygen abundance distribution (EDOD)
plotted in Fig.\,\ref{f:ddbb} with regard to
different [O/H]-[Fe/H] empirical relations,
deduced from the RB09 sample in presence of LTE
approximation via Eq.\,(\ref{eq:OF72}) (left
panel) and from the Fal09
sample in absence of LTE approximation, case SH1,
via Eq.\,(\ref{eq:OF70}) (right panel).
In the latter case,
the FC region has also been considered in
alternative to F and C regions separately.
For further details refer to the text.}
\label{t:figs}
\begin{center}
\begin{tabular}{lllll} \hline
\multicolumn{1}{c|}{}
& \multicolumn{2}{c|}{RB09}
& \multicolumn{2}{c}{Fal09 (SH1)} \\
\hline
\multicolumn{1}{l|}{transition} &
\multicolumn{1}{c}{$\phi$} &
\multicolumn{1}{c|}{$\psi$} &
\multicolumn{1}{c}{$\phi$} &
\multicolumn{1}{c}{$\psi$} \\
\hline
      &              &                 &              &              \\
O-A   & 9.4624 E$-$4 & $+$2.0383 E$-$3 & 3.1623 E$-$4 & $+$3.0083 E$-$1 \\ 
A-F   & 9.7001 E$-$3 & $+$9.9844 E$-$1 & 7.4271 E$-$3 & $+$8.6567 E$-$1 \\ 
F-C   & 6.1687 E$-$2 & $+$8.8953 E$-$1 & 1.2740 E$-$1 & $+$9.9952 E$-$1 \\ 
C-E   & 2.2854 E$-$1 & $-$3.3795 E$-$1 & 6.4083 E$-$1 & $-$9.5902 E$-$1 \\ 
E-O   & 7.1779 E$-$1 & $-$1.4421 E$-$0 & 3.1623 E$-$0 & $-$2.3121 E$-$0 \\
A-FC  &              &                 & 6.7114 E$-$3 & $+$8.0882 E$-$1 \\
FC-E  &              &                 & 6.2002 E$-$1 & $-$9.4785 E$-$1 \\
\hline       
\end{tabular}
\end{center} 
\end{table}  
A vertical line instead of a regression line
has been considered for the O region, where
no data exist.

In conclusion, the EDOD related to the inner
halo may be approximated, to a satisfactory
extent, as the sum of four (A, F, C, E) or
three (A, FC, E) regions, within which the
trend is linear.   An interpretations in
terms of simple models of chemical evolution
is highly attractive, as the corresponding
TDOD shows, in fact, a linear trend (Pagel
1989; Rocha-Pinto and Maciel 1996; C00; C01;
C07).

\section{The model}
\label{osm}

In their original formulation, simple
models of chemical evolution are
closed-box (CB) i.e. mass conservation
(gas + stars) holds (e.g., Searle and
Sargent 1972; Pagel and Patchett 1975).
In later formulation, a ``reservoir''
is added to the ``box'', where mass
conservation no longer occurs within
the box, but still holds within the
system box + reservoir.   Accordingly,
related models can be conceived as
closed-(box+reservoir) (CBR) models.
The gas
within the box is ``active'' in the
sense that allows star formation
according to a specified birth-rate
stellar function.   The gas within
the reservoir is ``inhibited'' in
the sense that no star formation
takes place.   Gas may outflow
from the box into the reservoir
(H76) or inflow into the box from
the reservoir  (C07).
The related TDOD is a straight
line where the slope (to the
knowledge of the author) is
negative in all cases studied in
literature (e.g., Pagel 1989;
Rocha-Pinto and Maciel 1996;
C00; C01; C07).

For this reason, the formulation
of CBR models shall be extended to
a TDOD with positive slope, under
the standard assumptions of CB
models: (i) instantaneous recycling
within the box, where stars are
divided into two categories, namely
(a) short-lived, which instantaneously
evolve, and (b) long-lived, all of
which are still evolving; (ii)
instantaneous mixing within the box,
where the gas returned from short-lived
stars is instantaneously mixed with the
interstellar medium yielding uniform
composition; and the standard assumptions
of CBR models: (iii) mass conservation
within the system (box+reservoir); (iv)
gas outflow from the box into the reservoir
or inflow into the box from the reservoir,
at a rate proportional to the star formation
rate; (v) inhibition of star formation
within the reservoir; (vi) gas outflow or
inflow with same composition with respect
to preexisting gas.

In this picture, the oxygen (or any other
primary element) mass fraction can be
determined, extending the procedure
followed for CB models (e.g., Pagel and
Patchett 1975; Wang and Silk 1993; C00)
to CBR models.

\subsection{Basic theory}
\label{bath}

The change in oxygen (or any other
primary element) gas mass, $M_{\rm gO}$,
is owing to four contributions, as:
\begin{equation}
\label{eq:dMgO}
\frac{\diff M_{\rm gO}}{\diff t}=\left(\frac{\diff M_{\rm gO}}{\diff t}
\right)_{\rm sf}+\left(\frac{\diff M_{\rm gO}}{\diff t}\right)_{\rm gr}+
\left(\frac{\diff M_{\rm gO}}{\diff t}\right)_{\rm sdu}+
\left(\frac{\diff M_{\rm gO}}{\diff t}\right)_{\rm sds}~~;
\end{equation}
related to subtraction via star formation
(sf), subtraction via outflow from the box
into the reservoir or addition via inflow
into the box from the reservoir (gr),
addition via unsynthesised gas from
short-lived stars (sdu), and addition via
newly synthesised gas from short-lived stars
(sds), respectively.

For simple CBR models, the following
relations hold:
\begin{lefteqnarray}
\label{eq:dMg}
&& \frac{\diff M_{\rm g}}{\diff t}=-\alpha\frac{\diff M_{\rm S}}{\diff t}
-\kappa\alpha\frac{\diff M_{\rm S}}{\diff t}=-\alpha(1+\kappa)
\frac{\diff M_{\rm S}}{\diff t}~~; \\
\label{eq:dMgO3}
&& \left(\frac{\diff M_{\rm gO}}{\diff t}\right)_{\rm sf}+
\left(\frac{\diff M_{\rm gO}}{\diff t}\right)_{\rm gr}+
\left(\frac{\diff M_{\rm gO}}{\diff t}\right)_{\rm sdu}=
-Z_{\rm O}\alpha(1+\kappa)\frac{\diff M_{\rm S}}{\diff t}~~; \\
\label{eq:dMgO4}
&& \left(\frac{\diff M_{\rm gO}}{\diff t}\right)_{\rm sds}=(1-Z)\hat{p}
\alpha\frac{\diff M_{\rm S}}{\diff t}~~;
\end{lefteqnarray}
where $M_{\rm g}$ is the mass in active gas,
$M_{\rm S}$ is the global mass in gas which
has been turned into stars, $M_{\rm gO}$ is
the oxygen mass in active gas, $\alpha$ is
the fraction in long-lived stars and stellar
remnants within a star generation (lock
parameter), $\kappa$ is the ratio of flowing
(outflow from the box into the reservoir or
inflow into the box from the reservoir) gas
rate to locking (in the form of long-lived
stars and stellar remnants) gas rate (flow
parameter), $Z_{\rm O}$ and $Z$ are the
oxygen and metal mass abundance, respectively,
within the active gas, and $\hat{p}$ is the
ratio of the oxygen mass newly synthesised
and returned to the interstellar medium,
for a metal-free initial composition%
\footnote{This detail is usually omitted in literature.},
to the mass locked up in long-lived stars
and stellar remnants (yield parameter).
The lock parameter, $\alpha$, the flow
parameter, $\kappa$, and the yield parameter,
$\hat{p}$, for sake of brevity, in the following
shall be quoted as the lock, the flow,
and the yield, respectively.

The definition of oxygen mass abundance,
$Z_{\rm O}$, fractional active gas mass,
$\mu$, and fractional star mass, $s$, read:
\begin{equation}
\label{eq:def}
Z_{\rm O}=\frac{M_{\rm gO}}{M_{\rm g}}~~;\quad\mu=\frac{M_{\rm g}}{M_0}~~;
\quad s=\alpha S=\frac{\alpha M_{\rm S}}{M_0}=\frac{M_{\rm s}}{M_0}~~;
\end{equation}
where $M_{\rm s}=\alpha M_{\rm S}$ is the
mass in long-lived stars and stellar remnants,
and $M_0$ is the total mass within the box
at the starting configuration.

The substitution of Eqs.\,(\ref{eq:dMg})-(\ref{eq:dMgO4})
into (\ref{eq:dMgO}) yields:
\begin{equation}
\label{eq:dMgO5}
\frac{\diff M_{\rm gO}}{\diff t}=\left[Z_{\rm O}-\frac{1-Z}{1+\kappa}\hat{p}
\right]\frac{\diff M_{\rm g}}{\diff t}~~;
\end{equation}
and the fractional oxygen gas mass,
via Eq.\,(\ref{eq:def}) reads:
\begin{equation}
\label{eq:fMgO}
\frac{M_{\rm gO}}{M_0}=\frac{M_{\rm gO}}{M_{\rm g}}
\frac{M_{\rm g}}{M_0}=Z_{\rm O}\mu~~;
\end{equation}
where, in addition:
\begin{equation}
\label{eq:dOmu}
\frac{\diff(Z_{\rm O}\mu)}{\diff t}=Z_{\rm O}\frac{\diff\mu}{\diff t}+
\mu\frac{\diff Z_{\rm O}}{\diff t}~~;
\end{equation}
on the other hand, in terms of fractional
masses, Eq.\,(\ref{eq:dMgO5}) may be cast
under the equivalent form:
\begin{equation}
\label{eq:dOmu2}
\frac{\diff(Z_{\rm O}\mu)}{\diff t}=\left[Z_{\rm O}-\frac{1-Z}{1+\kappa}
\hat{p}\right]\frac{\diff\mu}{\diff t}~~;
\end{equation}
and the combination of Eqs.\,(\ref{eq:dOmu})
and (\ref{eq:dOmu2}) yields:
\begin{lefteqnarray}
\label{eq:dOmu3}
&& \frac{\diff Z_{\rm O}}{1-Z}=-\hat{p}^\pprime\frac{\diff\mu}{\mu}~~; \\
\label{eq:ps}
&& \hat{p}^\pprime=\frac{\hat{p}}{1+\kappa}~~;
\end{lefteqnarray}
where $\hat{p}^\pprime$ is the effective
yield parameter, hereafter quoted as the
effective yield (H76).

In the special case of a linear $Z_{\rm O}$-$Z$
relation:
\begin{equation}
\label{eq:A}
\frac{1-Z_{\rm O}}{1-Z}=\frac1A~~;\qquad A<1~~;
\end{equation}
the combination of  Eqs.\,(\ref{eq:dOmu3})
and (\ref{eq:A}) produces:
\begin{equation}
\label{eq:dOmu4}
\frac{\diff Z_{\rm O}}{1-Z_{\rm O}}=-A\hat{p}^\pprime\frac{\diff\mu}{\mu}~~;
\end{equation}
which can be integrated.   After some
algebra, the result is:
\begin{equation}
\label{eq:ZOg}
Z_{\rm O}-(Z_{\rm O})_i=[1-(Z_{\rm O})_i]\left[1-\left(\frac{\mu}{\mu_i}
\right)^{A\hat{p}^\pprime}\right]~~;
\end{equation}
where the index, $i$, denotes the
starting configuration at the cosmic
time, $t_i$.   Reversing the role of
the variables, Eq.\,(\ref{eq:ZOg})
reads:
\begin{equation}
\label{eq:mug}
\frac{\mu}{\mu_i}=\left[\frac{1-Z_{\rm O}}{1-(Z_{\rm O})_i}\right]^{1/
(A\hat{p}^\pprime)}~~;
\end{equation}
which is monotonic in $Z_{\rm O}$
within the domain, $0\le Z_{\rm O}
\le1$.

Using the MacLaurin series development:
\begin{equation}
\label{eq:exps}
\left(\frac{\mu}{\mu_i}\right)^{A\hat{p}^\pprime}=\exp\left(A\hat{p}^\pprime
\ln\frac{\mu}{\mu_i}\right)=1+A\hat{p}^\pprime\ln\frac{\mu}{\mu_i}+...~~;
\end{equation}
under the further assumption that the
terms of higher order with respect to
the first can be neglected, Eq.\,(\ref{eq:ZOg})
reduces to:
\begin{lefteqnarray}
\label{eq:ZOl}
&& Z_{\rm O}-(Z_{\rm O})_i=-[1-(Z_{\rm O})_i]A\hat{p}^\pprime\ln\frac{\mu}
{\mu_i}~~; \\
\label{eq:disl}
&& \left\vert A\hat{p}^\pprime\ln\frac{\mu}{\mu_i}\right\vert<
\left\vert\hat{p}^\pprime\ln\frac{\mu}{\mu_i}\right\vert<
\left\vert\hat{p}^\pprime\ln\frac{\mu_f}{\mu_i}\right\vert\ll1~~;
\end{lefteqnarray}
where the index, $f$, denotes the ending
configuration at the cosmic time, $t_f$,
and $\mu_i<\mu<\mu_f$ or $\mu_i>\mu>\mu_f$
owing to Eq.\,(\ref{eq:mug}).   The initial
oxygen and metal abundance may safely be neglected
with respect to unity, $(Z_{\rm O})_i\le Z_i\ll1$,
which implies $A\to1$ via Eq.\,(\ref{eq:A}).
Accordingly, Eq.\,(\ref{eq:ZOl}) reduces to: 
\begin{equation}
\label{eq:ZOs}
Z_{\rm O}-(Z_{\rm O})_i=-\hat{p}^\pprime\ln\frac{\mu}{\mu_i}~~;
\end{equation}
which is the classical formulation (H76).

It is worth emphasyzing that the
inequality, expressed by Eq.\,(\ref{eq:disl}),
does not affect the instantaneous recycling
approximation, but only the general formulation
expressed by Eq.\,(\ref{eq:ZOg}) provided the
$Z_{\rm O}$-$Z$ relation is linear, according
to Eq.\,(\ref{eq:A}).   In general, it is assumed
the instantaneou recycling approximation holds
for sufficiently high fractional active gas
mass fraction, $\mu\appgeq0.1$ (e.g., Prantzos
2007, Fig.\,12 therein), but the threshold
could be lowered using Eq.\,(\ref{eq:ZOg})
instead of Eq.\,(\ref{eq:ZOl}).   In fact,
neglecting the terms of higher order in the
MacLaurin series, expressed by Eq.\,(\ref{eq:exps}),
makes oxygen abundance, $Z_{\rm O}$, increased
with respect to the general case, Eq.\,(\ref{eq:ZOg}),
which, in turn, is an upper limit with respect
to the exact formulation, Eq.\,(\ref{eq:dOmu3}).
More specifically, $\mu\to0$ implies $Z_{\rm O}
\to1$ according to Eq.\,(\ref{eq:ZOg}) instead
of $Z_{\rm O}\to Z_{\rm O}(\mu=0)<1$ due to the
presence of other metals.   Then the instantaneous
recycling approximation expressed by
Eq.\,(\ref{eq:ZOg}), or better by the integral
of Eq.\,(\ref{eq:dOmu3}), appears to hold well
even below the threshold, $\mu\approx0.1$.

In terms of fractional masses, Eq.\,(\ref{eq:dMg})
via Eq.\,(\ref{eq:def}) may be cast under the
equivalent form:
\begin{equation}
\label{eq:dmu}
\alpha(1+\kappa)\frac{\diff S}{\diff t}=(1+\kappa)\frac{\diff s}{\diff t}=
-\frac{\diff\mu}{\diff t}~~;
\end{equation}
which can be read in the following way:
any mass change in active gas is
counterbalanced by a change in long-lived
stars and stellar remnants plus a change
in gas outflow from the box into the reservoir
$(\kappa>0)$ or a change in gas inflow into
the box from the reservoir $(\kappa<0)$.
More specifically, a number of different
flow regimes may be distinguished as listed below.
\begin{description}
\item[$\bullet$\hspace{6.1mm}]
Outflow regime ($\kappa>0$), where star formation is
inhibited (H76).   For an exhaustive
description refer to earlier attempts (C00; C01).
\item[$\bullet$\hspace{6.1mm}]
Stagnation regime $(\kappa=0)$, where star formation is 
neither inhibited nor enhanced.   Accordingly,
CBR models reduce to CB models.   For an
exhaustive description refer to earlier attempts
(Searle and Sargent 1972; Pagel and Patchett
1975).
\item[$\bullet$\hspace{6.1mm}]
Moderate inflow regime ($-1<\kappa<0$), where star formation
is enhanced and active gas mass fraction
monotonically decreases in time.   For an
exhaustive description refer to an earlier
attempt (C07).
\item[$\bullet$\hspace{6.1mm}]
Steady inflow regime ($\kappa=-1$), where star formation is
enhanced and active gas mass fraction remains unchanged.
\item[$\bullet$\hspace{6.1mm}]
Strong inflow regime ($\kappa<-1$), where star formation is
enhanced and active gas mass fraction monotonically
increases in time.
\end{description}
The effective yield, $\hat{p}^\pprime$,
defined by Eq.\,(\ref{eq:ps}), cannot
exceed the real yield, $\hat{p}$, both
in outflow and in strong inflow regime,
while the contrary holds in moderate
inflow regime and, {\it a fortiori},
in steady inflow regime, where a
divergency occurs.   For this reason,
the effective yield cannot be considered
alone, but together with the factor,
$\ln[\mu/(\mu)_i]$.   More specifically,
the right-hand side of Eq.\,(\ref{eq:ZOs}),
$-\hat{p}^\pprime\ln[\mu/(\mu)_i]=-\hat{p}
\ln[\mu/(\mu)_i]/(1+\kappa)$, is positive
in all regimes due to the trend exhibited
by the active gas mass fraction.   In
steady inflow regime, the following relation
necessarily holds:
\begin{equation}
\label{eq:ZOs2}
\lim_{\kappa\to-1}\left(\frac{-\hat{p}}{1+\kappa}\ln\frac\mu{\mu_i}\right)=
\frac{Z_{\rm O}-(Z_{\rm O})_i}{(Z_{\rm O})_f-(Z_{\rm O})_i}=\frac{t-t_i}
{t_f-t_i}~~;
\end{equation}
in terms of the oxygen abundance related to
a constant active gas mass fraction, or in
terms of the cosmic time.

In any case, mass conservation follows from
integration of Eq.\,(\ref{eq:dmu}), as:
\begin{lefteqnarray}
\label{eq:masc}
&& \mu+s+D=\mu_0+s_0+D_0=\mu_0=1~~; \\
\label{eq:D}
&& D=\alpha\kappa S=\kappa s~~;
\end{lefteqnarray}
conformly to Eq.\,(\ref{eq:def}),
where $D$ is the gas mass fraction
which outflowed from the box into the
reservoir ($\kappa>0$) or inflowed
into the box from the reservoir
($\kappa<0$).
An equivalent form of
Eq.\,(\ref{eq:masc}) reads:
\begin{equation}
\label{eq:masc2}
\mu+s=1-D~~;
\end{equation}
where variables on the left and the right-hand
side member relate to the box and
the reservoir, respectively.

The birth-rate stellar function
(number of stars born per unit
volume, mass, and time), needs
to be specified for determining
the temporal behaviour of gas
and star fractional masses.
Following an earlier attempt
(C00), the selected choice
reads:
\begin{equation}
\label{eq:BRSF}
{\cal B}(\widetilde{m},\widetilde{t})=\frac BVf(\mu)\Phi(\widetilde{m})~~;
\quad\widetilde{m}=\frac m{m_\odot}~~;\quad\widetilde{t}=\frac t{\rm Gyr}~~;
\end{equation}
where $B$ is a normalization constant,
$V$ the volume of the box, $\Phi(\widetilde{m})$
the stellar initial mass function and
$f(\mu)$ an assigned function of the
active gas mass fraction, which is the
time-dependent term.   The number of
stars born per unit volume within an
infinitesimal dimensionless mass range,
$\widetilde{m}\mp\diff\widetilde{m}/2$,
and infinitesimal dimensionless time
range, $\widetilde{t}\mp\diff\widetilde
{t}/2$, is ${\cal B}(\widetilde{m},
\widetilde{t})\diff\widetilde{m}\diff
\widetilde{t}$.

The number of long-lived stars generated
(within the box) up to an assigned
dimensionless cosmic time, $\widetilde{t}$, is:
\begin{leftsubeqnarray}
\slabel{eq:Na}
&& N_{\rm\ell\ell}(\widetilde{t})=\int_{\widetilde{t}_i}^{\widetilde{t}}\diff
\widetilde{t}\int_{\widetilde{m}_{\rm mf}}^{\widetilde{m}_{\rm\ell\ell}}V
{\cal B}(\widetilde{m},\widetilde{t})\diff\widetilde{m}=BF(\widetilde{t})
I^\prime(5)~~; \\
\slabel{eq:Nb}
&& F(\widetilde{t})=\int_{\widetilde{t}_i}^{\widetilde{t}}f(\mu)\diff
\widetilde{t}~~;\quad I^\prime(5)=\int_{\widetilde{m}_{\rm mf}}^
{\widetilde{m}_{\rm\ell\ell}}\Phi(\widetilde{m})\diff\widetilde{m}~~;
\label{seq:N}
\end{leftsubeqnarray}
where $m_{\rm mf}$ is the lower stellar
mass limit, and $m_{\rm\ell\ell}$ the upper
mass limit of long-lived stars.

The mass fraction in stars globally
generated (within the box) up to a
dimensionless cosmic time, $\widetilde{t}$, is:
\begin{leftsubeqnarray}
\slabel{eq:Sa}
&& S(\widetilde{t})=\int_{\widetilde{t}_i}^{\widetilde{t}}\diff
\widetilde{t}\int_{\widetilde{m}_{\rm mf}}^{\widetilde{m}_{\rm Mf}}\frac V
{M_0}{\cal B}(\widetilde{m},\widetilde{t})\diff\widetilde{m}=CF(\widetilde{t})
~~; \\
\slabel{eq:Sb}
&& C=\frac{Bm_\odot}{M_0}I(1)~~;\quad I(1)=\int_{\widetilde{m}_{\rm mf}}^
{\widetilde{m}_{\rm Mf}}\widetilde{m}\Phi(\widetilde{m})\diff\widetilde{m}~~;
\label{seq:S}
\end{leftsubeqnarray}
where $m_{\rm Mf}$ is the upper stellar
mass limit.

The mass fraction in long-lived stars
and in gas outflowed from the box into
the reservoir or inflowed into the box
from the reservoir (in both cases with
equal composition with respect to the
preexisting gas), at an assigned
dimensionless cosmic time,
$\widetilde{t}$, are:
\begin{lefteqnarray}
\label{eq:s2}
&& s(\widetilde{t})=\alpha S(\widetilde{t})=\alpha CF(\widetilde{t})~~; \\
\label{eq:D2}
&& D(\widetilde{t})=\alpha\kappa S(\widetilde{t})=\alpha\kappa C
F(\widetilde{t})~~;
\end{lefteqnarray}
according to the definition of
lock and flow, respectively.

The star formation rate (within
the box), the lock rate, and the
flow rate at an assigned dimensionless
cosmic time, $\widetilde{t}$, are:
\begin{lefteqnarray}
\label{eq:dSt}
&& \frac{\diff S}{\diff\widetilde{t}}=Cf(\mu)~~; \\
\label{eq:dst}
&& \frac{\diff s}{\diff\widetilde{t}}=\alpha Cf(\mu)~~; \\
\label{eq:dDt}
&& \frac{\diff D}{\diff\widetilde{t}}=\alpha\kappa Cf(\mu)~~;
\end{lefteqnarray}
and the combination of 
Eqs.\,(\ref{eq:dmu}) and (\ref{eq:dSt})
yields:
\begin{equation}
\label{eq:dmu2}
\alpha(1+\kappa)Cf(\mu)=-\frac{\diff\mu}{\diff\widetilde{t}}~~;
\end{equation}
which is equivalent to:
\begin{equation}
\label{eq:mug2}
\int_{\mu_i}^\mu\frac{\diff\mu}{f(\mu)}=-\alpha(1+\kappa)C(\widetilde{t}-
\widetilde{t}_i)~~;
\end{equation}
in the special case, $f(\mu)=\mu$,
Eq.\,(\ref{eq:mug2}) can be integrated as:
\begin{equation}
\label{eq:fmu}
\mu=\mu_i\exp[-\alpha(1+\kappa)C(\widetilde{t}-\widetilde{t}_i)]~~;
\end{equation}
and the combination of
Eqs.\,(\ref{eq:ps}), (\ref{eq:ZOs}),
and (\ref{eq:fmu}) yields a linear
trend for the oxygen abundance:
\begin{equation}
\label{eq:Zfm1}
Z_{\rm O}-(Z_{\rm O})_i=\hat{p}\alpha C(\widetilde{t}-\widetilde{t}_i)~~;
\end{equation}
regardless of the flow, $\kappa$.

In the more general case, $f(\mu)=
\mu^\nu$, Eq.\,(\ref{eq:dSt})
reduces to a Schmidt (1959, 1963)
star formation law.
Leaving aside the above discussed
special case, $\nu=1$, Eq.\,(\ref
{eq:mug2}) can be integrated and,
after some algebra, the result is:
\begin{equation}
\label{eq:fmus}
\mu=\mu_i\left[1-\frac{1-\nu}{\mu_i^{1-\nu}}\alpha(1+\kappa)C(\widetilde{t}-
\widetilde{t}_i)\right]^{1/(1-\nu)}~~;\quad\nu\ne1~~;
\end{equation}
which, using the logarithm Taylor series,
$\ln(1+x)=x-x^2/2+x^3/3-...$, $\mid x\mid
<1$, and neglecting the terms of higher
order with respect to the first, may be
approximated as:
\begin{leftsubeqnarray}
\slabel{eq:fmuaa}
&& \mu=\mu_i\exp\left[-\mu_i^{\nu-1}\alpha(1+\kappa)C(\widetilde{t}-
\widetilde{t}_i)\right]~~; \\
\slabel{eq:fmuab}
&& \left\vert(1-\nu)\mu_i^{\nu-1}\alpha(1+\kappa)C(\widetilde{t}-
\widetilde{t}_i)\right\vert\ll1~~;
\label{seq:fmua}
\end{leftsubeqnarray}
in the limit, $\nu\to1$, 
Eqs.\,(\ref{eq:fmu}) and (\ref{eq:fmuaa})
coincide, as expected.   The combination
of Eqs.\,(\ref{eq:ps}), (\ref{eq:ZOs}),
and (\ref{seq:fmua}) yields:
\begin{equation}
\label{eq:Zfmus}
Z_{\rm O}-(Z_{\rm O})_i=\hat{p}\mu_i^{\nu-1}\alpha C(\widetilde{t}-
\widetilde{t}_i)~~;
\end{equation}
regardless of the flow, $\kappa$.
Then a Schmidt star formation law
with exponent, $\nu\approx1$, implies
a linear dependence of the oxygen
(or any other primary element)
abundance on the cosmic time.

In general, simple CBR models are described
by Eqs.\,(\ref{eq:ZOs}) and (\ref{eq:mug2}),
the latter only for including the temporal
behaviour.

\subsection{Theoretical differential oxygen
abundance distribution (TDOD)}
\label{TDOD}

For a selected spectral class, the
number of long-lived stars generated
(within the box) up to an assigned
dimensionless cosmic time, is:
\begin{leftsubeqnarray}
\slabel{eq:Nsca}
&& N(\widetilde{t})=\int_{\widetilde{t}_i}^{\widetilde{t}}\diff\widetilde{t}
\int_{\widetilde{m}_1}^{\widetilde{m}_2}V{\cal B}(\widetilde{m},\widetilde{t})
=C_{12}F(\widetilde{t})~~; \\
\slabel{eq:Nscb}
&& C_{12}=BI^\prime(1,2)~~;\quad I^\prime(1,2)=\int_{\widetilde{m}_1}^
{\widetilde{m}_2}\Phi(\widetilde{m})\diff\widetilde{m}~~;
\label{seq:Nsc}
\end{leftsubeqnarray}
where $m_1$ and
$m_2$ are the lower and upper mass
limit of the spectral class.

For simple CBR models, the oxygen
abundance is monotonically increasing
in time owing to Eqs.\,(\ref{eq:ZOg})
and (\ref{eq:fmus}), which implies the
number of stars born up to the cosmic
time, $t$, coincides with the number
of stars with oxygen abundance up to
the related value, $Z_{\rm O}(t)$, or
$N(t)=N(Z_{\rm O})$.   The same holds
for the long-lived star mass fraction
(including stellar remnants),
$s(t)=s(Z_{\rm O})$.   By use of
Eq.\,(\ref{eq:def}), the long-lived
star mass fraction and the number of
long-lived stars, $N_{\rm\ell\ell}$, can
be related as:
\begin{equation}
\label{eq:sNll}
\frac{s-s_i}{s_f-s_i}=\frac{\overline{m}N_{\rm\ell\ell}-\overline{m}(N_{\rm\ell\ell})_i}
{\overline{m}(N_{\rm\ell\ell})_f-\overline{m}(N_{\rm\ell\ell})_i}~~;
\end{equation}
where $\overline{m}$ is the mean mass
of long-lived stars (including stellar
remnants), $\overline{m}=M_s/N_{\rm\ell\ell}$.
Under the assumption
of a universal stellar initial mass
function, the fraction of long-lived
stars belonging to a selected spectral
class with respect to the total,
maintains unchanged, $N_{\rm\ell\ell}/N=$
const, and Eq.\,(\ref{eq:sNll})
translates into (Pagel and Patchett
1975):
\begin{equation}
\label{eq:sN}
\frac{s-s_i}{s_f-s_i}=\frac{N-N_i}{N_f-N_i}~~;
\end{equation}
and the combination of 
Eqs.\,(\ref{eq:masc}) and
(\ref{eq:D}) yields:
\begin{equation}
\label{eq:smu}
(1+\kappa)s=1-\mu~~;
\end{equation}
accordingly, Eq.\,(\ref{eq:sN})
may be cast under the equivalent
form:
\begin{equation}
\label{eq:Nmu}
\frac{N-N_i}{N_f-N_i}=\frac{1-\mu/\mu_i}{1-\mu_f/\mu_i}~~;
\end{equation}
where the ratio, $\mu/\mu_i$, is
expressed by Eq.\,(\ref{eq:ZOs})
with respect to the oxygen abundance,
$Z_{\rm O}$, and by Eqs.\,(\ref{eq:fmu})
and (\ref{eq:fmus}) with respect to
the dimensionless cosmic time,
$\widetilde{t}$.

In terms of the normalized oxygen
abundance, $\phi=Z_{\rm O}/(Z_
{\rm O})_\odot$, Eq.\,(\ref{eq:ZOs})
translates into:
\begin{leftsubeqnarray}
\slabel{eq:fimua}
&& \phi-\phi_i=-\frac{\hat{p}^\pprime}{(Z_{\rm O})_\odot}\ln\frac{\mu}{\mu_i}
~~; \\
\slabel{eq:fimub}
&& \left\vert\hat{p}^\pprime\ln\frac{\mu}{\mu_i}\right\vert<
\left\vert\hat{p}^\pprime\ln\frac{\mu_f}{\mu_i}\right\vert\ll1~~;
\label{seq:fimu}
\end{leftsubeqnarray}
and the derivation of 
Eqs.\,(\ref{eq:sN}) and
(\ref{eq:Nmu}) with respect
to $\phi$ yields:
\begin{equation}
\label{eq:dfi}
\frac{\diff N}{(N_f-N_i)\diff\phi}=\frac{\diff s}{(s_f-s_i)\diff\phi}=
\frac{-\diff(\mu/\mu_i)}{(1-\mu_f/\mu_i)\diff\phi}~~;
\end{equation}
which, using Eq.\,(\ref{seq:fimu}),
takes the explicit form:
\begin{equation}
\label{eq:dfie}
\frac{\diff N}{(N_f-N_i)\diff\phi}=\frac{\mu_i}{\mu_i-\mu_f}\frac
{(Z_{\rm O})_\odot}{\hat{p}^\pprime}\exp\left[-\frac{(Z_{\rm O})_\odot}
{\hat{p}^\pprime}(\phi-\phi_i)\right]~~;
\end{equation}
where the decimal logarithm of
the left-hand side is defined
as the TDOD (Pagel 1989; C00; C01):
\begin{lefteqnarray}
\label{eq:psit}
&& \psi(\phi)=\log\frac{\diff N}{(N_f-N_i)\diff\phi}=a\phi+b~~; \\
\label{eq:a}
&& a=-\frac1{\ln10}\frac{(Z_{\rm O})_\odot}{\hat{p}^\pprime}=
-\frac1{\ln10}\frac{(Z_{\rm O})_\odot}{\hat{p}}(1+\kappa)~~; \\
\label{eq:b}
&& b=\log\left[\frac{\mu_i}{\mu_i-\mu_f}(-\ln10)a\right]-a\phi_i~~;
\end{lefteqnarray}
which is represented by a straight
line on the $({\sf O}\phi\psi)$
plane.

The TDOD slope, $a$, defined by
Eq.\,(\ref{eq:a}), depends on the
flow regime discussed above.
More specifically, $a<0$ both in
outflow regime and in moderate
inflow regime; $a=0$ in steady
inflow regime; $a>0$ in strong
inflow regime.   The TDOD intercept,
$b$, defined by Eq.\,(\ref{eq:b}),
must necessarily fulfill the
condition, $\mu_f>0$, which implies
the inequality:
\begin{equation}
\label{eq:mu0}
b>b(\mu_f=0)=\log(-\ln10\,a)-a\phi_i~~;\quad a<0~~;
\end{equation}
on the other hand, $\mu_f>0$
directly follows from $a\ge0$.

\subsection{Fitting to empirical differential oxygen
abundance distribution (EDOD)}
\label{FEDO}

Both the EDOD and the TDOD can be
represented on the $({\sf O}\phi\psi)$
plane, but related normalizations are
different.   More specifically, the
former is normalized to the sample
population, $N$, according to
Eq.\,(\ref{eq:psie}), while the
latter is normalized to the computed
long-lived star population, $N_f-N_i$,
according to Eq.\,(\ref{eq:psit}).
Then the EDOD and the TDOD differ by
a normalization constant, $\log C_N$,
which must be taken into consideration
in performing the fitting procedure.
In other words, the TDOD, $\psi_
{\rm T}$, has to be vertically shifted
on the $({\sf O}\phi\psi)$ plane by
a value, $\log C_N$,
for matching to the EDOD, $\psi_
{\rm E}$.   The TDOD intercept,
defined by Eq.\,(\ref{eq:b}),
translates into:
\begin{equation}
\label{eq:bCN}
b=\log\left[\frac{C_{\rm N}\mu_i}{\mu_i-\mu_f}(-\ln10)a\right]-a\phi_i~~;
\end{equation}
where $a$ and $b$ can be
determined as the EDOD regression line
slope and intercept, respectively.

Using Eq.\,(\ref{eq:a}), the
empirical counterpart of
Eq.\,(\ref{eq:disl}), which
implies the validity of
Eq.\,(\ref{seq:fimu}), reads:
\begin{equation}
\label{eq:dise}
\left\vert\frac{(Z_{\rm O})_\odot}{\ln10}\frac1a\ln\frac\mu{\mu_i}\right\vert<
\left\vert\frac{(Z_{\rm O})_\odot}{\ln10}\frac1a\ln\frac{\mu_f}{\mu_i}\right
\vert\ll1~~;
\end{equation}
keeping in mind that the active
gas mass fraction, $\mu$,
monotonically changes as a
function of the
normalized oxygen abundance,
$\phi$, for simple CBR models.

The particularization of
Eq.\,(\ref{seq:fimu}) to the
final configuration, by use
of Eq.\,(\ref{eq:a}), reads: 
\begin{equation}
\label{eq:fimf}
\phi_f-\phi_i=\frac1{\ln10}\frac1a\ln\frac{\mu_f}{\mu_i}~~;
\end{equation}
and the combination of
Eqs.\,(\ref{eq:bCN})
and (\ref{eq:fimf}) yields:
\begin{equation}
\label{eq:CN}
C_{\rm N}=-\frac1{\ln10}\frac1a\exp_{10}(a\phi_i+b)\{1-\exp_{10}[a(\phi_f-
\phi_i)]\}~~;
\end{equation}
%
%
%
which may be determined from
the knowledge of the EDOD and
related regression line.

\subsection{Different stages of evolution}
\label{mscm}

The mere existence of a G-dwarf
problem in different regions of
the Galaxy (e.g., van den Bergh
1962; Schmidt 1963; H76;
Prantzos 2003; Ferreras et al.
2003) and perhaps in all galaxies
(Worthey et al. 1996; Henry and
Worthey 1999) implies the EDOD
cannot be fitted by a straight
line, as predicted by simple CBR
models, but by a continuous broken line at
most.   To this aim, simple CBR
models shall be extended by
allowing different flow regimes
during different stages of
evolution, and defined as simple
multistage closed-(box+reservoir)
(MCBR) models.   Accordingly,
the flow, $\kappa$, is different
in different stages, while the
equations of the model maintain
their formal expression where
variables and parameters are
indexed by a letter, U = I,
II, III, ..., which denotes the
stage under consideration.

With regard to the U-th stage,
the TDOD, defined by
Eq.\,(\ref{eq:psit}), reads:
\begin{equation}
\label{eq:psfiU}
\psi_{\rm U}(\phi_{\rm U})=\log\frac{(C_{\rm U})_{\rm N}\diff s_{\rm U}}
{[(s_{\rm U})_f-(s_{\rm U})_i]\diff\phi_{\rm U}}~~;
\end{equation}
where $(s_{\rm U})_i$ and $(s_{\rm U})_f$
are the fractional star mass at the
beginning and at the end, respectively,
of the U-th stage, and $\log (C_{\rm U})_{\rm N}$
is the related normalization constant
with respect to the EDOD, under the
assumption of a universal stellar
initial mass function, according to
Eq.\,(\ref{eq:sN}).   The combination
of Eqs.\,(\ref{eq:sN}), (\ref{eq:dfie}),
and (\ref{eq:psfiU}) yields:
\begin{lefteqnarray}
\label{eq:psiU}
&& \psi_{\rm U}(\phi_{\rm U})=a_{\rm U}\phi_{\rm U}+b_{\rm U}~~; \\
\label{eq:aU}
&& a_{\rm U}=-\frac1{\ln10}\frac{(Z_{\rm O})_\odot}{\hat{p}_{\rm U}^\pprime}=
-\frac1{\ln10}\frac{(Z_{\rm O})_\odot}{\hat{p}_{\rm U}}(1+\kappa_{\rm U})~~;
\\
\label{eq:bU}
&& b_{\rm U}=\log\left[\frac{(C_{\rm U})_{\rm N}(\mu_{\rm U})_i}{(\mu_
{\rm U})_i-(\mu_{\rm U})_f}(-\ln10)a_{\rm U}\right]-a_{\rm U}(\phi_{\rm U})_
i~~;
\end{lefteqnarray}
which is valid provided the following
inequality holds:
\begin{equation}
\label{eq:disfU}
\left\vert\frac{(Z_{\rm O})_\odot}{\ln10}\frac1{a_{\rm U}}\ln\frac
{\mu_{\rm U}}{(\mu_{\rm U})_i}\right\vert<
\left\vert\frac{(Z_{\rm O})_\odot}{\ln10}\frac1{a_{\rm U}}\ln\frac
{(\mu_{\rm U})_f}{(\mu_{\rm U})_i}\right\vert\ll1~~;
\end{equation}
in agreement with Eq.\,(\ref{eq:dise}).

The particularization of Eq.\,(\ref{eq:fimf})
to the U-th stage reads:
\begin{equation}
\label{eq:fimfU}
(\phi_{\rm U})_f-(\phi_{\rm U})_i=\frac1{\ln10}\frac1{a_{\rm U}}\ln\frac
{(\mu_{\rm U})_f}{(\mu_{\rm U})_i}~~;
\end{equation}
and the combination of
Eqs.\,(\ref{eq:aU}), (\ref{eq:bU}), and
(\ref{eq:fimfU}) yields:
\begin{equation}
\label{eq:CNU}
(C_{\rm U})_{\rm N}=-\frac1{\ln10}\frac1{a_{\rm U}}\exp_{10}[a_{\rm U}
(\phi_{\rm U})_i+b_{\rm U}]\{1-\exp_{10}\{a_{\rm U}[(\phi_{\rm U})_f-
(\phi_{\rm U})_i]\}\}~~;
\end{equation}
%
%
which may be determined from the knowledge
of the EDOD belonging to the
U-th stage and related regression
line.   In general, different samples
and/or different stages imply different
values of the normalization constant.

\subsection{Application to a special stellar system}
\label{feih}

For selected [O/H]-[Fe/H] relations,
the EDOD related to the fs10 sample
(taken as representative of ther inner Galactic
halo) can be divided into four or three
regions where the trend is linear to a
good extent, as shown in Figs.\,\ref
{f:ddbb} and \ref{f:tdbb}.   In the
light of the model, region A corresponds
to the first stage of evolution, where
the system is still assembling, characterized
by strong inflow regime ($\kappa<-1$, $a>0$);
region F corresponds to the second stage of
evolution, where the system is forming,
characterized by outflow regime ($\kappa>0$,
$a<0$); region C corresponds to the third
stage of evolution, where the system is
undergoing contraction, characterized by
outflow regime; region E corresponds to
the fourth step of evolution, where the
system has attained dynamical equilibrium,
characterized by outflow regime.

The gas and star mass fraction are left
unchanged passing from a selected stage
to the next one, which implies the validity
of the following relations:
\begin{leftsubeqnarray}
\slabel{eq:XUVa}
&& (X_{\rm U})_f=(X_{\rm V})_i~~;\quad(X_{\rm A})_i=X_i~~;\quad(X_{\rm E})_f=
X_f~~; \\
\slabel{eq:XUVb}
&& X=\mu,s,D~~;\quad{\rm U=A,F,C}~~;\quad{\rm V=F,C,E}~~;
\label{seq:XUV}
\end{leftsubeqnarray}
where $X_i$, $X_f$, are related
to the whole evolution regardless
of the stages.   Accordingly,
mass conservation during the U-th
stage may be expressed as:
\begin{equation}
\label{eq:mcU}
\mu_{\rm U}+(1+\kappa_{\rm U})s_{\rm U}=\mu_{\rm U}+s_{\rm U}+D_{\rm U}=
(\mu_{\rm U})_i+(s_{\rm U})_i+(D_{\rm U})_i~~;
\end{equation}
where the initial values are known via
Eq.\,(\ref{seq:XUV}).

The regression line in an assigned
region of the EDOD is defined by
the slope, $a_{\rm U}$, and the
intercept, $b_{\rm U}$, and the
intersections between regression
lines related to adjacent regions
mark initial and final values of
normalized oxygen abundance,
$(\phi_{\rm U})_i$ and $(\phi_
{\rm U})_f$.   The normalization
constant, $(C_{\rm U})_{\rm N}$,
is determined via Eq.\,(\ref{eq:CNU}).
The final value of the active gas
mass fraction, $(\mu_{\rm U})_f$,
follows from Eq.\,(\ref{eq:fimfU}).

The yield, $\hat{p}$, and the lock,
$\alpha$, depend on the birth-rate
stellar function.   For a power-law
stellar initial mass function, the
EDOD in different Galactic environments
is reproduced to a good extent from
simple CBR models with the following
output parameters (CM09):
\begin{leftsubeqnarray}
\slabel{eq:yfma}
&& \frac{\hat{p}}{(Z_{\rm O})_\odot}=1.0340~~; \\
\slabel{eq:yfmb}
&& \alpha_{2.9}=0.66604~~;\qquad\alpha_{2.35}=0.85360~~; \\
\slabel{eq:yfmc}
&& (\widetilde{m}_{\rm mf})_{2.9}=0.44449~~;\qquad
(\widetilde{m}_{\rm mf})_{2.35}=0.015436~~;
\label{seq:yfm}
\end{leftsubeqnarray}
for an assumed solar oxygen abundance,
$(Z_{\rm O})_\odot=0.0056$, where the
indices relate to
power-law exponents equal to 2.9 and 2.35,
respectively.
For further details refer to earlier
attempts (C01; C07).

The flow, $\kappa_{\rm U}$, is deduced
from Eq.\,(\ref{eq:aU}).   The final
star mass fraction, $(s_{\rm U})_f$,
and outflowed or inflowed gas mass
fraction, $(D_{\rm U})_f$, are determined
from Eq.\,(\ref{eq:mcU}) particularized
to the end of the U-th stage.

In the light of the model, the initial
star mass fraction, $s_i$, results from
the fractional mass in stars (including stellar
remnants) with normalized oxygen abundance,
$\phi<\phi_i$.   The above value, though
still uncertain at present, may be
understimated to a good extent as $s_i=0$.
In addition, $D_i=0$ without loss of
generality, which implies $\mu_i=1$.

The fractional mass of the box (with
respect to the initial value) attains
the maximum value at the end of A
stage, where gas inflows into the
box from the reservoir.   The related
value is $(\mu_{\rm A})_f+(s_{\rm A})_f$.
The fractional mass in stars at the end
of evolution, which coincides with the
end of E stage, is $(s_{\rm E})_f$.
Accordingly, the mass ratio of the box
at the end of evolution to the outflowed
gas reads:
\begin{equation}
\label{eq:MHB}
\frac{M_{\rm box}}{M_{\rm ofl}}=\frac{(\mu_{\rm E})_f+(s_{\rm E})_f}
{(\mu_{\rm A})_f+(s_{\rm A})_f-(\mu_{\rm E})_f-(s_{\rm E})_f}~~;
\end{equation}
provided the earlier stage A is in
inflow regime and the subsequent
stages F, C, (or FC), E, are in
outflow regime, as it is the case
for the EDOD and related TDOD
under consideration.

With the above values of input
parameters, 
the flow, the active gas mass
fraction, the star mass fraction,
and the outflowed or inflowed
gas mass fraction, at the end
of each stage of evolution,
can be computed.
The results are listed in Table
\ref{t:ralm} for the EDOD
related to the fs10 sample for
[O/H]-[Fe/H] empirical relations
deduced from the RB09 sample
(top panel) and Fal09 sample,
case SH1 (middle and bottom panel),
see Fig.\,\ref{f:ddbb}.
\begin{table}
\caption{Input parameters (deduced
from the regression lines), $(\phi_{\rm U})_i$,
$a_{\rm U}$, $b_{\rm U}$, and output
parameters, $\kappa_{\rm U}$, $(\mu_{\rm U})_f$,
$(s_{\rm U})_f$, $(D_{\rm U})_f$, for
simple MCBR models fitting to the EDOD
related to the fs10 sample, for 
[O/H]-[Fe/H] empirical relations
deduced from the RB09 sample
(top panel) and Fal09 sample,
case SH1 (middle and bottom panel).
Four (A, F, C, E; top and middle
panels) or three (A, FC, E; bottom
panel) stages of evolution are
considered, according to the linear
trends exhibited by the EDOD
(Fig.\,\ref{f:ddbb}).   Stages O
before A and after E are not
considered as no sample object
lies within the corresponding
metallicity range.   Other input
parameters are $(Z_{\rm O})_\odot=
0.0056$; $\hat{p}/(Z_{\rm O})_\odot=
1.0340$; $\mu_i=(\mu_{\rm O})_f=1$;
$s_i=(s_{\rm O})_f=0$; $D_i=(D_
{\rm O})_f=0$; where the index, O,
denotes oxygen with regard to the
solar abundance and stage O of
evolution otherwise.   For further
details refer to the text.}
\label{t:ralm}
\begin{center}
\begin{tabular}{llllllll} \hline
\multicolumn{1}{l}{U} &
\multicolumn{1}{c}{$(\phi_{\rm U})_i$} &
\multicolumn{1}{c}{$a_{\rm U}$} &
\multicolumn{1}{c}{$b_{\rm U}$} &
\multicolumn{1}{c}{$\kappa_{\rm U}$} &
\multicolumn{1}{c}{$(\mu_{\rm U})_f$} &
\multicolumn{1}{c}{$(s_{\rm U})_f$} &
\multicolumn{1}{c}{$(D_{\rm U})_f$} \\
\hline

      &             &                &                &                &             &             &                \\
A     & 9.4624E$-$4 & $+$1.1382E$+$2 & $-$1.0567E$-$1 & $-$2.7201E$+$2 & 9.9175E$-$0 & 3.2905E$-$2 & $-$8.9504E$-$0 \\
F     & 9.7001E$-$3 & $-$2.0950E$+$0 & $+$1.0188E$-$0 & $+$3.9881E$+$0 & 7.7178E$-$0 & 4.7391E$-$1 & $-$7.1917E$-$0 \\
C     & 6.1687E$-$2 & $-$7.3565E$+$0 & $+$1.3433E$-$0 & $+$1.6515E$+$1 & 4.5710E$-$1 & 8.8844E$-$1 & $-$3.4554E$-$1 \\
E     & 2.2854E$-$1 & $-$2.2569E$+$0 & $+$1.7784E$-$1 & $+$4.3735E$+$0 & 3.5961E$-$2 & 9.6681E$-$1 & $-$2.7737E$-$3 \\
O     & 7.1779E$-$1 &                &                &                &             &             &                \\
      &             &                &                &                &             &             &                \\
A     & 3.1623E$-$4 & $+$7.9432E$+$1 & $+$2.7571E$-$1 & $-$1.9012E$+$2 & 3.6714E$-$0 & 1.4125E$-$2 & $-$2.6855E$-$0 \\
F     & 7.4271E$-$3 & $-$3.8855E$+$0 & $+$8.9453E$-$1 & $+$8.2913E$+$0 & 1.2551E$-$0 & 2.7531E$-$1 & $-$5.3043E$-$1 \\
C     & 1.2740E$-$1 & $-$2.6460E$+$0 & $+$7.3661E$-$1 & $+$5.3000E$+$0 & 5.4973E$-$2 & 4.6581E$-$1 & $+$4.7922E$-$1 \\
E     & 6.4083E$-$1 & $-$5.3661E$-$1 & $-$6.1514E$-$1 & $+$2.7764E$-$1 & 2.4384E$-$3 & 5.0693E$-$1 & $+$4.9063E$-$1 \\
O     & 3.1623E$-$0 &                &                &                &             &             &                \\
      &             &                &                &                &             &             &                \\
A     & 3.1623E$-$4 & $+$7.9432E$+$1 & $+$2.5751E$-$1 & $-$1.9012E$+$2 & 3.2209E$-$0 & 1.1743E$-$2 & $-$2.2327E$-$0 \\
FC    & 6.7114E$-$3 & $-$2.8643E$+$0 & $+$8.2804E$-$1 & $+$5.8196E$+$0 & 5.6405E$-$2 & 4.7577E$-$1 & $+$4.6782E$-$1 \\
E     & 6.2002E$-$1 & $-$5.3661E$-$1 & $-$6.1514E$-$1 & $+$4.7764E$-$1 & 2.4384E$-$3 & 5.1801E$-$1 & $+$4.7955E$-$1 \\
O     & 3.1623E$-$0 &                &                &                &             &             &                \\
\hline                            
\end{tabular}                     
\end{center}                      
\end{table}                       
With regard to the bottom panel,
stages F and C are merged into
a single one, FC.

In the former case (top panel),
where the empirical [O/H]-[Fe/H]
relation has been determined
using the LTE approximation
(RB09), the mass of the box is
increased by a factor of about
10 at the end of A stage, and
is reduced to about the initial
value at the end of E stage.
The mass ratio, $M_{\rm box}/
M_{\rm ofl}$, is about one tenth
via Eq.\,(\ref{eq:MHB}).

In the latter case (middle and
bottom panel), 
where the empirical [O/H]-[Fe/H]
relation has been determined
relaxing the LTE approximation
(Fal09), the mass of the box is
increased by a factor of about
3-4 at the end of A stage, and
is reduced to about one half
the initial
value at the end of E stage.
The mass ratio, $M_{\rm box}/
M_{\rm ofl}$, is about one sixth
via Eq.\,(\ref{eq:MHB}).

In any case, the initial mass of
the box at the beginning of A stage
(proto-inner halo) is comparable to
the mass of the box at the end of E
stage (present inner halo).

\subsection{Discussion} \label{disc}

Though TDODs calculated using simple
MCBR models of chemical evolution fit
to EDODs with different linear trends
in different regions, as shown in
Fig.\,\ref{f:tdbb},
still the application to the inner
Galactic halo remains speculative in
absence of further improvement.  The
main reasons are outlined below.

First, the EDOD has been derived from
a fictitious sample, fs10, which, in
turn, has been deduced from two incomplete
samples, RN91 and H\,V, both biased
(to a different extent)
towards low metallicities and affected
by disk contamination.   A homogeneous
and unbiased sample would be needed to
this respect.    The EDOD high-metallicity
tail could be largely due to disk
contamination as stage E begins at
[Fe/H$]\approx-1$, which is close to both the
transition from halo to bulge/disk
globular cluster (e.g., Mackey and
van den Bergh 2005) and the peak of
bulge EDOD (Sadler et al. 1996).
Different linear trends in different
regions are also shown by 
globular clusters (C07), bulge (C07),
and disk (Caimmi 2008) EDOD.

A linear trend exhibited by the EDOD
related to different Galactic environments
or galactic environments provided the
G-dwarf problem is universal (Worthey
et al. 1996; Henry and Worthey 1999),
implies outflowing or inflowing gas
with same composition as the preexisting
gas.   If the above mentioned effect is
real, the chemical evolution within the
building blocks of galactic subsystems,
in particular the inner Galactic halo,
occurred to a similar extent.

Second, using different methods yields
different [O/H]-[Fe/H] empirical relations,
as shown in Fig.\,\ref{f:FeHOH}, even in
dealing with coinciding sample objects,
as listed in Table \ref{t:rbfa}.   With
regard to the inner Galactic halo, the
initial mass deduced using two selected
[O/H]-[Fe/H] empirical relations, has
been found to change by a factor of about
two.   Future improvement would imply
direct oxygen abundance determinations
(Ramirez et al. 2007) where different
methods yield consistent results or,
at least, a [O/H]-[Fe/H] empirical
relation which attains general consensus.

Third, though the assumption of null
star mass fraction holds to a
good extent for the starting configuration,
still the presence of stars
with lower metallicity with respect to
sample objects, [Fe/H$]<-4.2$, has to be
considered.   This extremely low metallicity
has currently been detected in about a
dozen of stars down to [Fe/H$]\approx-5.4$
(e.g., Beers and Christlieb 2005).
Basing on theoretical arguments, a lower
limit oxygen abundance in Pop II stars
has been determined as [O/H$]=-3.05\mp0.20$
(Bromm and Loeb 2003), where earlier
nucleosynthesis comes from more massive
Pop III stars.

If the transition from the latter to the
former population was not instantaneous,
and coeval Pop III and Pop II stars were
generated (Smith et al. 2009), then the
stellar initial mass function can no
longer be considered as universal in time
for Pop II stars.   Accordingly, model
evolution must be started after the last
Pop III star has undergone supernova
explosion, which has been assumed to
take place when the metal abundance is
[Fe/H$]=-4.2$.   In absence (to the
knowledge of the author) of a reliable
estimate of the mass fraction in stars
and stellar remnants with initial metal
abundance, [Fe/H$]<-4.2$, a null value
has been assigned.

For a stellar system resembling the inner
Galactic halo, a nontrivial question is
concerned with
the nature of the reservoir where gas
outflows after the assembling (A) stage.
A natural candidate is the bulge, as
suggested by comparison between specific
angular momentum distributions related to
halo and bulge stars (Wyse and Gilmore
1992).   According to recent attempts,
(1) the Galactic halo is made of an inner
component which exhibits a modest
prograde rotation and a metallicity peak
at [Fe/H] $\approx-1.6$, and an outer
component which
exhibits a clear retrograde net rotation
and a metallicity peak at [Fe/H] $\approx-2.2$
(Carollo et al. 2007, 2010); and (2) the Galactic
bulge is made of an inner component which
exhibits bar-like kinematics and metal-rich
population, and an outer component which exhibits
spheroid kinematics and metal-poor population
(Babusiaux et al. 2010).  Then a
similar specific angular momentum distribution
related to inner halo and outer bulge is
expected.

In this view, the mass ratio of the box at
the end of evolution to the gas outflowed
from the box into the reservoir during the
evolution, $M_{\rm box}/M_{\rm ofl}$, expressed
by Eq.\,(\ref{eq:MHB}), should reproduce the
mass ratio of (inner) halo to (outer) bulge, as:
\begin{equation}
\label{eq:HBr}
\frac{(M_{\rm H})_{\rm inn}}{(M_{\rm B})_{\rm out}}=
\frac{(M_{\rm H})_{\rm inn}/M_{\rm H}}{(M_{\rm B})_{\rm out}/M_{\rm B}}
\frac{M_{\rm H}}{M_{\rm B}}~~;
\end{equation}
which is equivalent to:
\begin{equation}
\label{eq:HBr2}
\frac{(M_{\rm H})_{\rm inn}}{M_{\rm H}}=\frac{(M_{\rm H})_{\rm inn}/
{(M_{\rm B})_{\rm out}}}{M_{\rm H}/M_{\rm B}} 
\frac{(M_{\rm B})_{\rm out}}{M_{\rm B}}~~;
\end{equation}
where the indices, inn and out, denote
the inner and the outer component,
respectively, H the halo, and B the
bulge.   The the inner
halo to the outer bulge mass ratio, deduced from
the results listed in Table \ref{t:ralm},
is $(M_{\rm H})_{\rm inn}/(M_{\rm B})_
{\rm out}\approx0.10$-0.20.   The halo to
bulge mass ratio is currently estimated as
$M_{\rm H}/M_{\rm B}\approx0.05$-0.10.
Then simple MCBR models considered here
yield an inner halo fractional
mass (normalized to the halo) which is
comparable with, or exceeding by a factor
up to 4, the outer bulge fractional
mass (normalized to the bulge).
On the other hand, quantitative results
cannot be expected
for the above mentioned reasons.   In
conclusion, the (inner) halo to (outer)
bulge mass ratio appears to be an additional
output parameter provided by simple MCBR
models of chemical
evolution, with regard to the inner Galactic
halo.

In the light of the model, the inner Galactic
halo can be considered independently of the
outer halo and the bulge for the following
reasons.   The outer halo is both dynamically
and chemically decoupled from the inner halo
(Carollo et al. 2010), which implies accretion
by smaller subunits (mainly dwarf spheroidal
galaxies) after the formation of the inner halo.
In fact, globular clusters belonging to the
inner and outer halo are usually classified
as ``old halo'' and ``young halo'', respectively
(e.g., Mackey and van den Bergh 2005).
Similarly, a (inner) halo - (metal-poor) bulge collapse
can be inferred by comparison of related
specific angular momentum distributions
(Wise and Gilmore 1992), which implies bulge
formation is subsequent to inner halo formation,
and then the (metal-poor) bulge may be considered
as a reservoir for the outflowing gas, in the
sense specified by the model.

\section{Conclusion}
\label{conc}

Under the assumption that two samples
of halo stars, RN91 and H\,V, are
equally representative of the inner
Galactic halo within the metallicity
range, $-3.0\le[$Fe/H$]\le-2.8$, a
fictitious sample, fs10, has been
built up and taken as representative
of the inner
Galactic halo within the metallicity
range, $-4.2\le[$Fe/H$]\le+0.2$.
The related differential empirical
oxygen abundance distribution (EDOD)
has been established using different
[O/H]-[Fe/H] empirical relations,
deduced from different samples (RB09,
Fal09, Sal09) where different methods
have been exploited for determining
the oxygen abundance.

More precisely, the EDOD has been
deduced from the fs10 sample by
use of two alternative [O/H]-[Fe/H]
empirical relations: one, determined
from the RB09 sample in presence of
local thermodynamical equilibrium
(LTE) approximation (RB09), and one
other, determined from the Fal09
sample in absence of LTE approximation
with due account taken of the inelastic
collisions via neutral H atoms, SH1
(Fal09, $S_{\rm H}=1$), as shown in
Fig.\,\ref{f:FeHOH}.

A linear trend has been exhibited by
related EDODs within three or four
regions, as shown in Fig.\,\ref{f:ddbb},
and the slope and intercept estimators
of corresponding regression lines have
been determined together with their
dispersions, as listed in Table \ref
{t:inte} using different interpolation
methods.   It has been pointed out that
the earlier trend, characterized by
positive slope, is a signature of a
G-dwarf problem.

The main uncertainties on the EDOD
have been recognized as related to (1)
biases on the RN91 and H\,V samples
due to selection effects towards
sufficiently low ([Fe/H$]<-4.2$) and
sufficiently high ([Fe/H$]>-2.8$)
metal abundance (H\,V) and disk
contamination for [Fe/H$]>-2.0$
(RN91; H\,V), and (2) lack of clear
indications on a recommended method
for determining oxygen abundance,
as shown in Table \ref{t:rbfa} and
in Fig.\,\ref{f:rbfa} for 11 stars
in common among the RB09 and the
Fal09 sample.

Fitting a theoretical differential
oxygen abundance distribution (TDOD)
to the above discussed EDOD has needed
an extension of simple closed-box (CB)
models of chemical evolution on two
respects.   First, the system has been
conceived as made of a box and a reservoir,
where the following processes have been
allowed: gas outflow from the box into the
reservoir (H76), moderate
gas inflow (C07), steady and strong
gas inflow (current paper) into the
box from the reservoir.   Simple
closed-(box+reservoir) (CBR) models have
exhibited mass conservation within the
system (box+reservoir), while it has been
violated within a single subsystem (box
or reservoir).   Second, the history of
the system has been conceived as a
succession of different stages characterized
by different outflow or inflow rate.
Simple multistage closed-(box+reservoir)
(MCBR) models have been found to yield
TDODs in the form of continuous broken
lines, which can fit to related
EDODs.

An application of MCBR models to a
special stellar system resembling the
inner Galactic halo has been made with
fiducial values of input parameters
which cannot be deduced from the EDOD.
The metal abundance at the beginning
and at the end of each stage, have
been inferred from the intersection
of regression lines fitting to adjacent
regions of the EDOD, as shown in
Fig.\,\ref{f:tdbb}.   The mass ratio
of the box at the end of evolution to
the gas outflowed from the box into
the reservoir through the evolution,
has been determined as
$M_{\rm box}/M_{\rm ofl}\approx0.11$
for a [O/H]-[Fe/H] empirical relation
deduced from the RB09 sample in presence
of LTE approximation (RB09), and
$M_{\rm box}/M_{\rm ofl}\approx0.16$-0.19
for a [O/H]-[Fe/H] empirical relation
deduced from the Fal09 sample in absence
of LTE approximation 
with due account taken of the inelastic
collisions via neutral H atoms, SH1
(Fal09, $S_{\rm H}=1$), with regard
to four and three stages of evolution,
respectively.

For current estimates of the halo-to-bulge
mass ratio, $M_{\rm halo}/M_{\rm bulge}
\approx0.05$-0.10, the inner halo fractional
mass (normalized to the halo) has been
found to be comparable with, or exceeding
by a factor up to 4, the metal-poor bulge
fractional mass (normalized to the bulge).
On the other hand, it has been considered
that quantitative predictions cannot be
made for the Galaxy unless complete and
unbiased samples of the inner Galactic
halo are available, and
discrepancies among [O/H]-[Fe/H] empirical
relations related to different samples
and different methods are removed.


\appendix
\section*{Appendix}

\section{Tables}\label{a:tab}

Tables \ref{t:FeHOHp}, \ref{t:psph}, and \ref{t:ralm} 
may be exceedingly large in the text, and for this
reason they are broken to be completely accessible.

%
\begin{table*}
\caption[par]{Table \ref{t:FeHOHp}, left.}
\begin{center}
\begin{tabular}{llllll}
\multicolumn{2}{c|}{}
& \multicolumn{4}{c}{[O/H] = 0.72 [Fe/H]\hspace{10mm}(RB09)}
\\
\hline\noalign{\smallskip}
\multicolumn{1}{c}{$B_{\rm F}^-$} &
\multicolumn{1}{c}{$B_{\rm F}^+$} &
\multicolumn{1}{c}{$B_{\rm O}^-$} &
\multicolumn{1}{c}{$B_{\rm O}^+$} &
\multicolumn{1}{c}{$\phi$} &
\multicolumn{1}{c}{$\Delta^\mp\phi$} 
\\
\noalign{\smallskip}
\hline\noalign{\smallskip}
$-$4.2 & $-$4.0 & $-$3.024 & $-$2.880 & 1.1322D$-$3 & 1.8601D$-$4  \\
$-$4.0 & $-$3.8 & $-$2.880 & $-$2.736 & 1.5774D$-$3 & 2.5914D$-$4  \\
$-$3.8 & $-$3.6 & $-$2.736 & $-$2.592 & 2.1976D$-$3 & 3.6102D$-$4  \\
$-$3.6 & $-$3.4 & $-$2.592 & $-$2.448 & 3.0615D$-$3 & 5.0296D$-$4  \\
$-$3.4 & $-$3.2 & $-$2.448 & $-$2.304 & 4.2652D$-$3 & 7.0071D$-$4  \\
$-$3.2 & $-$3.0 & $-$2.304 & $-$2.160 & 5.9421D$-$3 & 9.7619D$-$4  \\
$-$3.0 & $-$2.8 & $-$2.160 & $-$2.016 & 8.2783D$-$3 & 1.3600D$-$3  \\
$-$2.8 & $-$2.6 & $-$2.016 & $-$1.872 & 1.1533D$-$2 & 1.8947D$-$3  \\
$-$2.6 & $-$2.4 & $-$1.872 & $-$1.728 & 1.6067D$-$2 & 2.6396D$-$3  \\
$-$2.4 & $-$2.2 & $-$1.728 & $-$1.584 & 2.2384D$-$2 & 3.6774D$-$3  \\
$-$2.2 & $-$2.0 & $-$1.584 & $-$1.440 & 3.1185D$-$2 & 5.1231D$-$3  \\
$-$2.0 & $-$1.8 & $-$1.440 & $-$1.296 & 4.3445D$-$2 & 7.1373D$-$3  \\
$-$1.8 & $-$1.6 & $-$1.296 & $-$1.152 & 6.0526D$-$2 & 9.9434D$-$3  \\
$-$1.6 & $-$1.4 & $-$1.152 & $-$1.008 & 8.4322D$-$2 & 1.3853D$-$2  \\
$-$1.4 & $-$1.2 & $-$1.008 & $-$0.864 & 1.1747D$-$1 & 1.9299D$-$2  \\
$-$1.2 & $-$1.0 & $-$0.864 & $-$0.720 & 1.6366D$-$1 & 2.6887D$-$2  \\
$-$1.0 & $-$0.8 & $-$0.720 & $-$0.576 & 2.2800D$-$1 & 3.7457D$-$2  \\
$-$0.8 & $-$0.6 & $-$0.576 & $-$0.432 & 3.1764D$-$1 & 5.2184D$-$2  \\
$-$0.6 & $-$0.4 & $-$0.432 & $-$0.288 & 4.4253D$-$1 & 7.2700D$-$2  \\
$-$0.4 & $-$0.2 & $-$0.288 & $-$0.144 & 6.1651D$-$1 & 1.0128D$-$1  \\
$-$0.2 & $+$0.0 & $-$0.144 & $+$0.000 & 8.5890D$-$1 & 1.4110D$-$1  \\
$+$0.0 & $+$0.2 & $+$0.000 & $+$0.144 & 1.1966D$-$0 & 1.9658D$-$1  \\
$+$0.2 & $+$0.4 & $+$0.144 & $+$0.288 & 1.6670D$-$0 & 2.7386D$-$1  \\
\noalign{\smallskip}
\hline              
\end{tabular}       
\end{center}        
\end{table*}
\begin{table*}
\caption[par]{Table \ref{t:FeHOHp}, right.}
\begin{center}
\begin{tabular}{llllll}
\multicolumn{2}{c|}{}
& \multicolumn{4}{c}{[O/H] = [Fe/H] + 0.70\hspace{10mm}(Fal09)} \\
\hline\noalign{\smallskip}
\multicolumn{1}{c}{$B_{\rm F}^-$} &
\multicolumn{1}{c}{$B_{\rm F}^+$} &
\multicolumn{1}{c}{$B_{\rm O}^-$} &
\multicolumn{1}{c}{$B_{\rm O}^+$} &
\multicolumn{1}{c}{$\phi$} &
\multicolumn{1}{c}{$\Delta^\mp\phi$} 
\\
\noalign{\smallskip}
\hline\noalign{\smallskip}
$-$4.2 & $-$4.0 & $-$3.5 & $-$3.3 & 4.0871D$-$4 & 9.2480D$-$5 \\
$-$4.0 & $-$3.8 & $-$3.3 & $-$3.1 & 6.4776D$-$4 & 1.4657D$-$4 \\
$-$3.8 & $-$3.6 & $-$3.1 & $-$2.9 & 1.0266D$-$3 & 2.3230D$-$4 \\
$-$3.6 & $-$3.4 & $-$2.9 & $-$2.7 & 1.6271D$-$3 & 3.6817D$-$4 \\
$-$3.4 & $-$3.2 & $-$2.7 & $-$2.5 & 2.5788D$-$3 & 5.8351D$-$4 \\
$-$3.2 & $-$3.0 & $-$2.5 & $-$2.3 & 4.0871D$-$3 & 9.2480D$-$4 \\
$-$3.0 & $-$2.8 & $-$2.3 & $-$2.1 & 6.4776D$-$3 & 1.4657D$-$3 \\
$-$2.8 & $-$2.6 & $-$2.1 & $-$1.9 & 1.0266D$-$2 & 2.3230D$-$3 \\
$-$2.6 & $-$2.4 & $-$1.9 & $-$1.7 & 1.6271D$-$2 & 3.6817D$-$3 \\
$-$2.4 & $-$2.2 & $-$1.7 & $-$1.5 & 2.5788D$-$2 & 5.8351D$-$3 \\
$-$2.2 & $-$2.0 & $-$1.5 & $-$1.3 & 4.0871D$-$2 & 9.2480D$-$3 \\
$-$2.0 & $-$1.8 & $-$1.3 & $-$1.1 & 6.4776D$-$2 & 1.4657D$-$2 \\
$-$1.8 & $-$1.6 & $-$1.1 & $-$0.9 & 1.0266D$-$1 & 2.3230D$-$2 \\
$-$1.6 & $-$1.4 & $-$0.9 & $-$0.7 & 1.6271D$-$1 & 3.6817D$-$2 \\
$-$1.4 & $-$1.2 & $-$0.7 & $-$0.5 & 2.5788D$-$1 & 5.8351D$-$2 \\
$-$1.2 & $-$1.0 & $-$0.5 & $-$0.3 & 4.0871D$-$1 & 9.2480D$-$2 \\
$-$1.0 & $-$0.8 & $-$0.3 & $-$0.1 & 6.4776D$-$1 & 1.4657D$-$1 \\
$-$0.8 & $-$0.6 & $-$0.1 & $+$0.1 & 1.0266D$+$0 & 2.3230D$-$1 \\
$-$0.6 & $-$0.4 & $+$0.1 & $+$0.3 & 1.6271D$+$0 & 3.6817D$-$1 \\
$-$0.4 & $-$0.2 & $+$0.3 & $+$0.5 & 2.5788D$+$0 & 5.8351D$-$1 \\
$-$0.2 & $+$0.0 & $+$0.5 & $+$0.7 & 4.0871D$+$0 & 9.2480D$-$1 \\
$+$0.0 & $+$0.2 & $+$0.7 & $+$0.9 & 6.4776D$+$0 & 1.4657D$-$0 \\
$+$0.2 & $+$0.4 & $+$0.9 & $+$1.1 & 1.0266D$+$1 & 2.3230D$-$0 \\
\noalign{\smallskip}
\hline              
\end{tabular}       
\end{center}        
\end{table*}
\begin{table*}
\caption[par]{Table \ref{t:psph}, left.}
\begin{center}
\begin{tabular}{llllrrr}
\multicolumn{4}{c|}{[O/H] = 0.72 [Fe/H]\hspace{10mm}(RB09)}
&\multicolumn{3}{c}{$\Delta N$} \\
\hline\noalign{\smallskip}
\multicolumn{1}{c}{$\phi$} & \multicolumn{1}{c}{$\phantom{0}\psi$} &
\multicolumn{1}{c}{$\Delta^-\psi$} & \multicolumn{1}{c}{$\Delta^+\psi$} &
\multicolumn{1}{c}{fs10}  & 
\multicolumn{1}{c}{H\,V}  & 
\multicolumn{1}{c}{RN91}  
\\
\noalign{\smallskip}
\hline\noalign{\smallskip}                                                                                           
1.1322D$-$3 & $-$1.4181D$-$1 & 5.3329D$-$1 & 2.3226D$-$1 &    2 &   2 &  0 \\
1.5774D$-$3 &                &             &             &    0 &   0 &  0 \\
2.1976D$-$3 &                &             &             &    0 &   0 &  1 \\
3.0615D$-$3 & $+$2.0434D$-$1 & 1.4793D$-$1 & 1.1014D$-$1 &   12 &  12 &  1 \\
4.2652D$-$3 & $+$6.3437D$-$1 & 4.7712D$-$1 & 2.2185D$-$1 &   45 &  45 &  2 \\
5.9421D$-$3 & $+$7.0048D$-$1 & 3.2187D$-$1 & 1.8282D$-$1 &   73 &  73 &  2 \\
8.2783D$-$3 & $+$8.9728D$-$1 & 1.8947D$-$1 & 1.3148D$-$1 &  160 & 160 &  8 \\
1.1533D$-$2 & $+$9.6413D$-$1 & 1.4107D$-$1 & 1.0631D$-$1 &  260 & 198 & 13 \\
1.6067D$-$2 & $+$1.0284D$-$0 & 1.0691D$-$1 & 8.5725D$-$2 &  420 & 281 & 21 \\
2.2384D$-$2 & $+$9.4240D$-$1 & 9.9155D$-$2 & 8.0671D$-$2 &  480 & 337 & 24 \\
3.1185D$-$2 & $+$9.4967D$-$1 & 8.1707D$-$2 & 6.8742D$-$2 &  680 & 399 & 34 \\
4.3445D$-$2 & $+$9.1764D$-$1 & 7.0967D$-$2 & 6.0983D$-$2 &  880 & 313 & 44 \\
6.0526D$-$2 & $+$9.2258D$-$1 & 5.8986D$-$2 & 5.1924D$-$2 & 1240 & 231 & 62 \\
8.4322D$-$2 & $+$6.9376D$-$1 & 6.5516D$-$2 & 5.6916D$-$2 & 1020 & 229 & 51 \\
1.1747D$-$1 & $+$4.7566D$-$1 & 7.1860D$-$2 & 6.1640D$-$2 &  860 & 209 & 43 \\
1.6366D$-$1 & $+$1.4535D$-$1 & 9.0970D$-$2 & 7.5175D$-$2 &  560 & 308 & 28 \\
2.2800D$-$1 & $-$3.3187D$-$1 & 1.4107D$-$1 & 1.0631D$-$1 &  260 & 268 & 13 \\
3.1764D$-$1 & $-$5.1063D$-$1 & 1.4793D$-$1 & 1.1014D$-$1 &  240 & 178 & 12 \\
4.4253D$-$1 & $-$8.8871D$-$1 & 2.0618D$-$1 & 1.3924D$-$1 &  140 & 109 &  7 \\
6.1651D$-$1 & $-$1.1788D$-$0 & 2.5744D$-$1 & 1.6053D$-$1 &  100 &  45 &  5 \\
8.5890D$-$1 &                &             &             &    0 &  33 &  0 \\
1.1966D$-$0 & $-$2.1658D$-$0 & $+\infty$   & 3.0103D$-$1 &   20 &   3 &  1 \\
1.6670D$-$0 &                &             &             &    0 &   6 &  0 \\
\noalign{\smallskip}
\hline                                                      
\end{tabular}                                               
\end{center}
\end{table*}
\begin{table*}
\caption[par]{Table \ref{t:psph}, right.}
\begin{center}
\begin{tabular}{llllrrr}
\multicolumn{4}{c|}{[O/H] = [Fe/H] + 0.70\hspace{10mm}(Fal09)}
&\multicolumn{3}{c}{$\Delta N$} \\
\hline\noalign{\smallskip}
\multicolumn{1}{c}{$\phi$} & 
\multicolumn{1}{c}{$\phantom{0}\psi$} &
\multicolumn{1}{c}{$\Delta^-\psi$} & \multicolumn{1}{c}{$\Delta^+\psi$} &
\multicolumn{1}{c}{fs10}  & 
\multicolumn{1}{c}{H\,V}  & 
\multicolumn{1}{c}{RN91}  \\
\noalign{\smallskip}
\hline\noalign{\smallskip}                                                                                           
4.0871D$-$4 & $+$1.6168D$-$1 & 5.3329D$-$1 & 2.3226D$-$1 &    2 &   2 &  0 \\
6.4776D$-$4 &                &             &             &    0 &   0 &  0 \\
1.0266D$-$3 &                &             &             &    0 &   0 &  1 \\
1.6271D$-$3 & $+$3.3983D$-$1 & 1.4793D$-$1 & 1.1014D$-$1 &   12 &  12 &  1 \\
2.5788D$-$3 & $+$7.1386D$-$1 & 4.7712D$-$1 & 2.2185D$-$1 &   45 &  45 &  2 \\
4.0871D$-$3 & $+$7.2397D$-$1 & 3.2187D$-$1 & 1.8282D$-$1 &   73 &  73 &  2 \\
6.4776D$-$3 & $+$8.6477D$-$1 & 1.8947D$-$1 & 1.3148D$-$1 &  160 & 160 &  8 \\
1.0266D$-$2 & $+$8.7562D$-$1 & 1.4107D$-$1 & 1.0631D$-$1 &  260 & 198 & 13 \\
1.6271D$-$2 & $+$8.8390D$-$1 & 1.0691D$-$1 & 8.5725D$-$2 &  420 & 281 & 21 \\
2.5788D$-$2 & $+$7.4189D$-$1 & 9.9155D$-$2 & 8.0671D$-$2 &  480 & 337 & 24 \\
4.0871D$-$2 & $+$6.9316D$-$1 & 8.1707D$-$2 & 6.8742D$-$2 &  680 & 399 & 34 \\
6.4776D$-$2 & $+$6.0513D$-$1 & 7.0967D$-$2 & 6.0983D$-$2 &  880 & 313 & 44 \\
1.0266D$-$1 & $+$5.5407D$-$1 & 5.8986D$-$2 & 5.1924D$-$2 & 1240 & 231 & 62 \\
1.6271D$-$1 & $+$2.6925D$-$1 & 6.5516D$-$2 & 5.6916D$-$2 & 1020 & 229 & 51 \\
2.5788D$-$1 & $-$4.8510D$-$3 & 7.1860D$-$2 & 6.1640D$-$2 &  860 & 209 & 43 \\
4.0871D$-$1 & $-$3.9116D$-$1 & 9.0970D$-$2 & 7.5175D$-$2 &  560 & 308 & 28 \\
6.4776D$-$1 & $-$9.2438D$-$1 & 1.4107D$-$1 & 1.0631D$-$1 &  260 & 268 & 13 \\
1.0266D$+$0 & $-$1.1591D$-$0 & 1.4793D$-$1 & 1.1014D$-$1 &  240 & 178 & 12 \\
1.6271D$+$0 & $-$1.5932D$-$0 & 2.0618D$-$1 & 1.3924D$-$1 &  140 & 109 &  7 \\
2.5788D$+$0 & $-$1.9393D$-$0 & 2.5744D$-$1 & 1.6053D$-$1 &  100 &  45 &  5 \\
4.0871D$+$0 &                &             &             &    0 &  33 &  0 \\
6.4776D$+$0 & $-$3.0383D$-$0 & $+\infty$   & 3.0103D$-$1 &   20 &   3 &  1 \\
1.0266D$+$1 &                &             &             &    0 &   6 &  0 \\
\noalign{\smallskip}
\hline                                                      
\end{tabular}                                               
\end{center}                                                
\end{table*}
\begin{table}
\caption{Table \ref{t:ralm}, left.}
\begin{center}
\begin{tabular}{llll} \hline
\multicolumn{1}{l}{U} &
\multicolumn{1}{c}{$(\phi_{\rm U})_i$} &
\multicolumn{1}{c}{$a_{\rm U}$} &
\multicolumn{1}{c}{$b_{\rm U}$} \\
\hline

   &             &                &                \\
A  & 9.4624E$-$4 & $+$1.1382E$+$2 & $-$1.0567E$-$1 \\
F  & 9.7001E$-$3 & $-$2.0950E$+$0 & $+$1.0188E$-$0 \\
C  & 6.1687E$-$2 & $-$7.3565E$+$0 & $+$1.3433E$-$0 \\
E  & 2.2854E$-$1 & $-$2.2569E$+$0 & $+$1.7784E$-$1 \\
O  & 7.1779E$-$1 &                &                \\
                 &                &                \\
A  & 3.1623E$-$4 & $+$7.9432E$+$1 & $+$2.7571E$-$1 \\
F  & 7.4271E$-$3 & $-$3.8855E$+$0 & $+$8.9453E$-$1 \\
C  & 1.2740E$-$1 & $-$2.6460E$+$0 & $+$7.3661E$-$1 \\
E  & 6.4083E$-$1 & $-$5.3661E$-$1 & $-$6.1514E$-$1 \\
O  & 3.1623E$-$0 &                &                \\
                 &                &                \\
A  & 3.1623E$-$4 & $+$7.9432E$+$1 & $+$2.5751E$-$1 \\
FC & 6.7114E$-$3 & $-$2.8643E$+$0 & $+$8.2804E$-$1 \\
E  & 6.2002E$-$1 & $-$5.3661E$-$1 & $-$6.1514E$-$1 \\
O  & 3.1623E$-$0 &                &                \\
\hline                            
\end{tabular}                     
\end{center}                      
\end{table}                       
\begin{table}
\caption{Table \ref{t:ralm}, right.}
\begin{center}
\begin{tabular}{lllll} \hline
\multicolumn{1}{l}{U} &
\multicolumn{1}{c}{$\kappa_{\rm U}$} &
\multicolumn{1}{c}{$(\mu_{\rm U})_f$} &
\multicolumn{1}{c}{$(s_{\rm U})_f$} &
\multicolumn{1}{c}{$(D_{\rm U})_f$} \\
\hline

                    &             &             &                \\
A  & $-$2.7201E$+$2 & 9.9175E$-$0 & 3.2905E$-$2 & $-$8.9504E$-$0 \\
F  & $+$3.9881E$+$0 & 7.7178E$-$0 & 4.7391E$-$1 & $-$7.1917E$-$0 \\
C  & $+$1.6515E$+$1 & 4.5710E$-$1 & 8.8844E$-$1 & $-$3.4554E$-$1 \\
E  & $+$4.3735E$+$0 & 3.5961E$-$2 & 9.6681E$-$1 & $-$2.7737E$-$3 \\
O  &                &             &             &                \\
                    &             &             &                \\
A  & $-$1.9012E$+$2 & 3.6714E$-$0 & 1.4125E$-$2 & $-$2.6855E$-$0 \\
F  & $+$8.2913E$+$0 & 1.2551E$-$0 & 2.7531E$-$1 & $-$5.3043E$-$1 \\
C  & $+$5.3000E$+$0 & 5.4973E$-$2 & 4.6581E$-$1 & $+$4.7922E$-$1 \\
E  & $+$2.7764E$-$1 & 2.4384E$-$3 & 5.0693E$-$1 & $+$4.9063E$-$1 \\
O  &                &             &             &                \\
                    &             &             &                \\
A  & $-$1.9012E$+$2 & 3.2209E$-$0 & 1.1743E$-$2 & $-$2.2327E$-$0 \\
FC & $+$5.8196E$+$0 & 5.6405E$-$2 & 4.7577E$-$1 & $+$4.6782E$-$1 \\
E  & $+$4.7764E$-$1 & 2.4384E$-$3 & 5.1801E$-$1 & $+$4.7955E$-$1 \\
O  &                &             &             &                \\
\hline                            
\end{tabular}                     
\end{center}                      
\end{table}                       

\end{document}